\documentclass[preprint]{elsarticle}

\usepackage[
  a4paper, mag=1000, includefoot,
  left=3cm, right=2cm, top=2cm, bottom=2cm, headsep=1cm, footskip=1cm
]{geometry}

\journal{Neural Networks}

\usepackage[english]{babel}
\usepackage{csquotes}

\usepackage{amsmath}
\usepackage[colorlinks=true]{hyperref}

\graphicspath{{./fig/}}

\renewcommand{\Re}{\mathop{\mathrm{Re}}}

\begin{document}

\begin{frontmatter}

\title{Discovering dynamical features of Hodgkin-Huxley-type model of
  physiological neuron using artificial neural network}

\author{Pavel V. Kuptsov\corref{cor1}}
\ead{kupav@mail.ru}

\author{Nataliya V. Stankevich}
\ead{stankevichnv@mail.ru}

\author{Elmira R. Bagautdinova}
\ead{bagautdinovaer@mail.ru}

\cortext[cor1]{Corresponding author}

\address{Laboratory of topological methods in dynamics, HSE University, 25/12
  Bolshaya Pecherskaya str., Nizhny Novgorod 603155, Russia}

\begin{abstract}
  We consider Hodgkin-Huxley-type model that is a stiff ODE system with two fast
  and one slow variables. For the parameter ranges under consideration the
  original version of the model has unstable fixed point and the oscillating
  attractor that demonstrates bifurcation from bursting to spiking
  dynamics. Also a modified version is considered where the bistability occurs
  such that an area in the parameter space appears where the fixed point becomes
  stable and coexists with the bursting attractor. For these two systems we
  create artificial neural networks that are able to reproduce their
  dynamics. The created networks operate as recurrent maps and are trained on
  trajectory cuts sampled at random parameter values within a certain
  range. Although the networks are trained only on oscillatory trajectory cuts,
  it also discover the fixed point of the considered systems. The position and
  even the eigenvalues coincide very well with the fixed point of the initial
  ODEs. For the bistable model it means that the network being trained only on
  one brunch of the solutions recovers another brunch without seeing it during
  the training. These results, as we see it, are able to trigger the development
  of new approaches to complex dynamics reconstruction and discovering. From the
  practical point of view reproducing dynamics with the neural network can be
  considered as a sort of alternative method of numerical modeling intended for
  use with contemporary parallel hard- and software.
\end{abstract}


\end{frontmatter}

\section{Introduction}

Application of artificial neural networks in nonlinear dynamics in some aspects
is a well developed area. Basically this is related to system modeling, state
space reconstruction and time series forecasting. A comprehensive survey is
beyond the scope of our paper, however some interesting topics, both early and
contemporary, can be mentioned: adaptive system modeling and control using
neural networks~\cite{Levin1991185}, reconstruction of the El Ni\~{n}o attractor
with neural networks~\cite{GrigLatif94}, using neural networks for combined
state space reconstruction and forecasting~\cite{BolNak2000}, reconstruction of
chaotic time series by neural networks~\cite{Tronci2003}, reconstructing a
dynamical system and forecasting with deep neural
networks~\cite{Wang2021}. However in view of the current success
of deep learning in various areas, the use of machine learning methods for
complex dynamics analysis can be extended. The promising perspective is related
with generalization ability of the networks. If a network is trained on data
generated by a dynamical system there are reasons to expect that it will extract
the essence of the trained data. The subsequent evaluation of the network will
result in discovering new features otherwise unnoticed. Another promising
perspective is related with the fact that the neural networks can be treated as
universal approximators~\cite{Kolmogorov56,Kolmogorov57,Arnold57,
  Cybenko1989,Haykin2009}.  A neural network can be trained to recover almost
any, even very complicated, functional dependence. In particular it means that
the networks can be used to model dynamical systems. Along with the
generalization properties it potentially can open interesting perspectives in
new approaches state space reconstruction and forecasting as well as revealing
new dynamical properties. From the practical point of view reproducing dynamics
with the neural network can be considered as a sort of alternative method of
numerical modeling intended for use with contemporary parallel hard- and
software. In particular it is suitable for so called AI accelerators, a hardware
dedicated to deal with artificial neural
networks~\cite{TPU2017,FutureAI,Karras2020}.

In this paper we focus on dynamics modeling using neural networks. Previously we
considered a simple two layer network as a recurrent map that is able to model
various dynamical systems including Lorenz system, R\"{o}ssler system and also
the Hindmarsh–Rose model~\cite{NNDyn21}. For these three examples the created
neural network map demonstrated high quality of reconstruction including
recovery of the Lyapunov spectra. However the dynamics of the well known
Hodgkin-Huxley-type model~\cite{Sherman1988} with two fast and one slow
variables is found to require a more sophisticated approach. In this paper we
create and analyze an artificial neural network that is able to reproduce this
dynamics. For the considered parameter ranges the system has unstable fixed
point and oscillating attractor that can be either bursting or spiking. Also a
modification to the Hodgkin-Huxley-type model is considered that introduces
bistability: an area in the parameter space appears where the fixed point
becomes stable and coexists with the bursting attractor. The created network
operates as a recurrent map, i.e., it reproduces the dynamics as a discrete time
system. This is trained as a feedforward network using standard back propagation
routine on trajectory cuts sampled at random parameter values within a certain
range. The network structure is developed to take into account the different
time scales of the variables of the modeled system. Although the network is
trained only on oscillatory trajectory cuts, the resulting recurrent map also
discovers the fixed point whose position and even the eigenvalues coincide well
with the fixed point of the initial ODE system. In particular it means that when
a bistable regime is modeled, the network being trained on one brunch of
solutions discovers another one, never seen in the course of training.

Before embarking on our study we need to mention another interesting approach to
modeling dynamics using the machine learning methods. In
papers~\cite{Kong2021a,Kong2021b} chaotic dynamics is recovered using so called
reservoir computing system. This system operates as a recurrent map and its
training is similar to the training of the recurrent neural network. The
training dataset comprises a set of long trajectories computed at various
parameter values and the training is done on them at once. The reservoir
computing system includes high dimensional iterations of the hidden layer
vector. The iterations include a multiplication by a non-trainable random matrix
and application of a nonlinear function. The modeled dynamical variable after
multiplication by also non-trainable random input matrix is added to the hidden
layer vector before each iteration, the iteration is done and then a value of
the dynamical variable at the next time step is computed as a result of
multiplication of the obtained hidden layer vector by an output matrix. It is
the only matrix whose elements are tuned in course of the training. The approach
based on reservoir computing demonstrates the successful prediction of critical
transition~\cite{Kong2021a}, reveals of the emergence of transient
chaos~\cite{Kong2021b}, allows performing time series
analysis~\cite{ThronJung22}.

The difference of our approach is that the state space dimension of our neural
network model is exactly the same as for the modeled dynamics. It does not
include any random non-trainable matrices. The resulting recurrent map is
written in an explicit form and admits its analysis: we are able to find
analytical expressions for the Jacobian matrix and analyze its eigen values.

The paper is organized as follows. First, in Sec.~\ref{sec:model_eqs} we discuss
the modeled system, then in Sec.~\ref{sec:netw} we introduce the neural network
model and finally in Sec.~\ref{sec:dynam} the dynamics of the neural network
model is analyzed. In Sec.~\ref{sec:concl} the obtained results are outlined.

\section{\label{sec:model_eqs}Modeled system}

We consider the simplified pancreatic beta-cell model based on the
Hodgkin-Huxley formalism~\cite{Sherman1988}. In addition to the original system
its modified version is also considered that includes bistability as suggested
by Stankevich and Mosekilde~\cite{StanMos2017}. This modification is introduced
to demonstrate that an additional voltage-dependent potassium current that is
activated in the region around the original, unstable equilibrium point results
in the coexistence of a stable equilibrium point with a state of continuous
bursting.
\begin{equation}\label{eq:sys}
  \begin{aligned}
    \tau \dot V &= -I_{Ca}(V) - I_{K}(V,n)-I_{K2}(V)-I_S(V,S),\\
    \tau \dot n &= \sigma\,[n_{\infty}(V)-n],\\
    \tau_{S} \dot S &= S_{\infty}(V)-S.
  \end{aligned}
\end{equation}
Here, $V$ represents the membrane potential, $n$ may be interpreted as the
opening probability of the potassium channels, and $S$ accounts for the presence
of a slow variable in the system. The variables $I_{Ca}(V)$ and $I_{K}(V,n)$ are
the calcium and potassium currents, $g_{Ca}$ and $g_K$ are the associated
conductances, and $V_{Ca}$ and $V_K$ are the respective Nernst (or reversal)
potentials. Together with $I_{S}(V, S)$, the slow calcium current $I_{Ca}$ and
the potassium current $I_{K}$ define the three transmembrane currents,
Eqs.~\eqref{eq:origsys_ica}, \eqref{eq:origsys_ik}, and
\eqref{eq:origsys_is}. The gating variables for $m$, $n$, and $S$ represent the
opening probabilities of the fast and slow potassium channels,
Eq.~\eqref{eq:orig_gates}. The modification resulting in the stabilization of
the equilibrium point is introduced as the voltage-dependent potassium current
that varies strongly with the membrane potential right near this equilibrium
point so that its stabilization can occur without affecting the global flow in
the model.  The suggested form of the potassium current is specified by
Eq.~\eqref{eq:modif_cur} where the function~\eqref{eq:modif_prob} represents the
opening probability for the introduced new type of potassium channel.
\begin{gather}
  I_{Ca}(V)=g_{Ca} \, m_{\infty}(V) \, (V-V_{Ca}), \label{eq:origsys_ica}\\
  I_{K}(V,n)=g_{K}\, n\, (V-V_{K}), \label{eq:origsys_ik}\\
  I_{S}(V,S)=g_{S}\, S\, (V-V_{K}), \label{eq:origsys_is}\\
  I_{K2}(V)=g_{K2} \, p_{\infty}(V) \, (V-V_{K}), \label{eq:modif_cur}\\
  \omega_{\infty}(V) = \left(
    1+\exp \frac{V_{\infty}-V}{\theta_{\omega}}
  \right)^{-1}\hspace{-1.2em},\hspace{1.2em} \omega=m,n,S, \label{eq:orig_gates}\\
  p_{\infty}(V) = \left(
    \exp \frac{V-V_{p}}{\theta_{p}}+
    \exp \frac{V_{p}-V}{\theta_{p}}
  \right)^{-1}. \label{eq:modif_prob}
\end{gather}

\begin{table*}
  \caption{\label{tab:param}Numerical values of parameters of the model~\eqref{eq:sys}}
  \begin{center}
  \begin{tabular}{p{0.2\textwidth}p{0.2\textwidth}p{0.2\textwidth}p{0.2\textwidth}}
    \hline
    $\tau=0.02\,\text{s}$      & $\tau_S=35\,\text{s}$       & $\sigma=0.93$              & {} \\
    $g_{Ca}=3.6$               & $g_{K}=10$                  & $g_{S}=4$                  & $g_{K2}=0.12$ \\
    $V_{Ca}=25\,\text{mV}$     & $V_{K}=-75\,\text{mV}$      & {}                         & {} \\
    $\theta_{m}=12\,\text{mV}$ & $\theta_{n}=5.6\,\text{mV}$ & $\theta_{S}=10\,\text{mV}$ & $\theta_{p} = 1\,\text{mV}$ \\
    $V_{m}=-20\,\text{mV}$     & $V_n=-16\,\text{mV}$        & $V_{S}=-36\,\text{mV}$     & $V_{p}=-49.5\,\text{mV}$ \\
    \hline
  \end{tabular}
  \end{center}
\end{table*}

\begin{figure}
  \begin{center}
    \begin{tabular}{cc}
      \includegraphics[width=0.48\columnwidth]{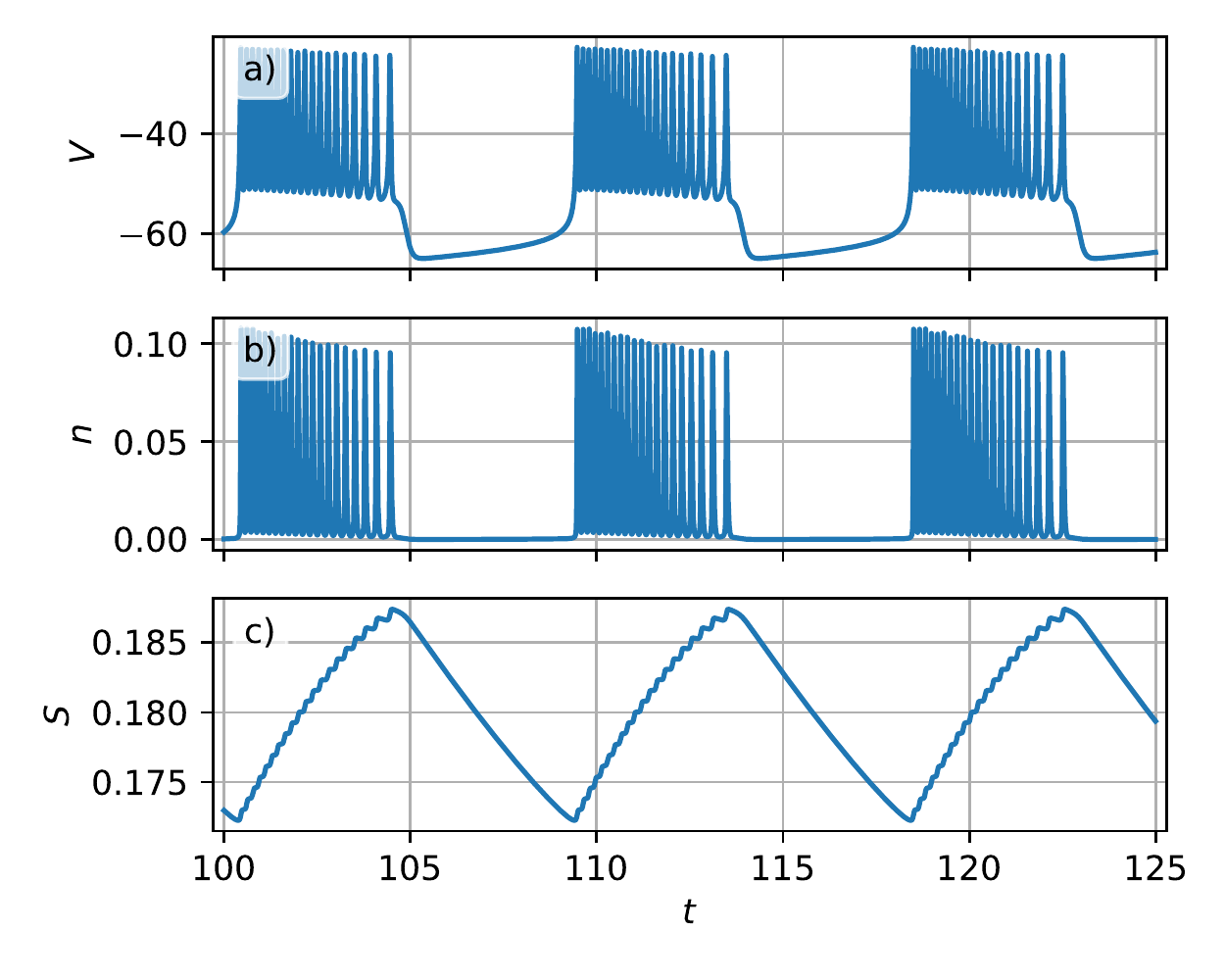} &
      \includegraphics[width=0.48\columnwidth]{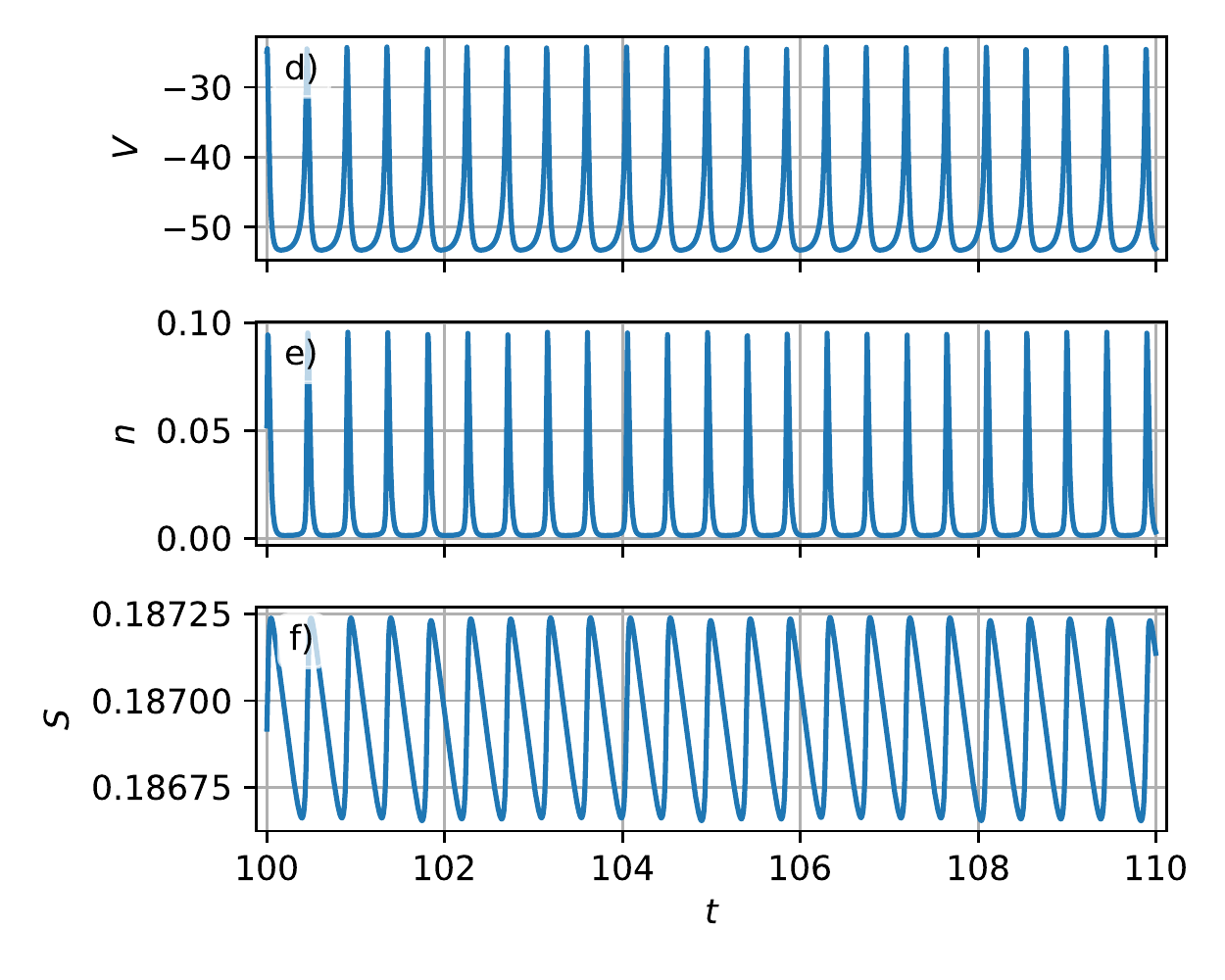}
    \end{tabular}
  \end{center}
  \caption{\label{fig:tser}Time series of the original system~\eqref{eq:sys},
    i.e., at $g_{K2}=0$. Panels (a,b,c): bursting oscillations at
    $V_S=-36$. Panels (d,e,f) spiking regime at $V_S=-33$. Other parameters see
    in Tab.~\ref{tab:param}.}
\end{figure}

\begin{figure}
  \begin{center}
    \begin{tabular}{cc}
      \includegraphics[width=0.48\columnwidth]{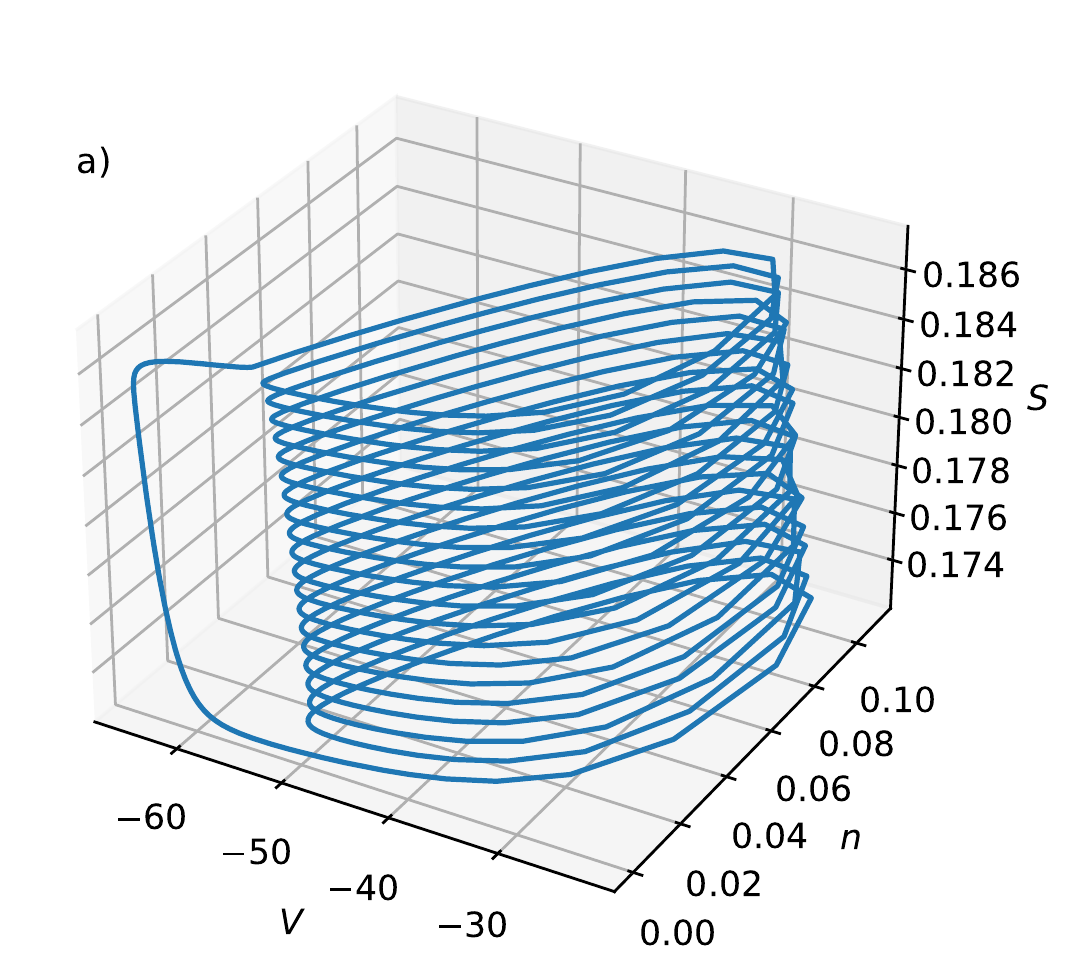} &
      \includegraphics[width=0.48\columnwidth]{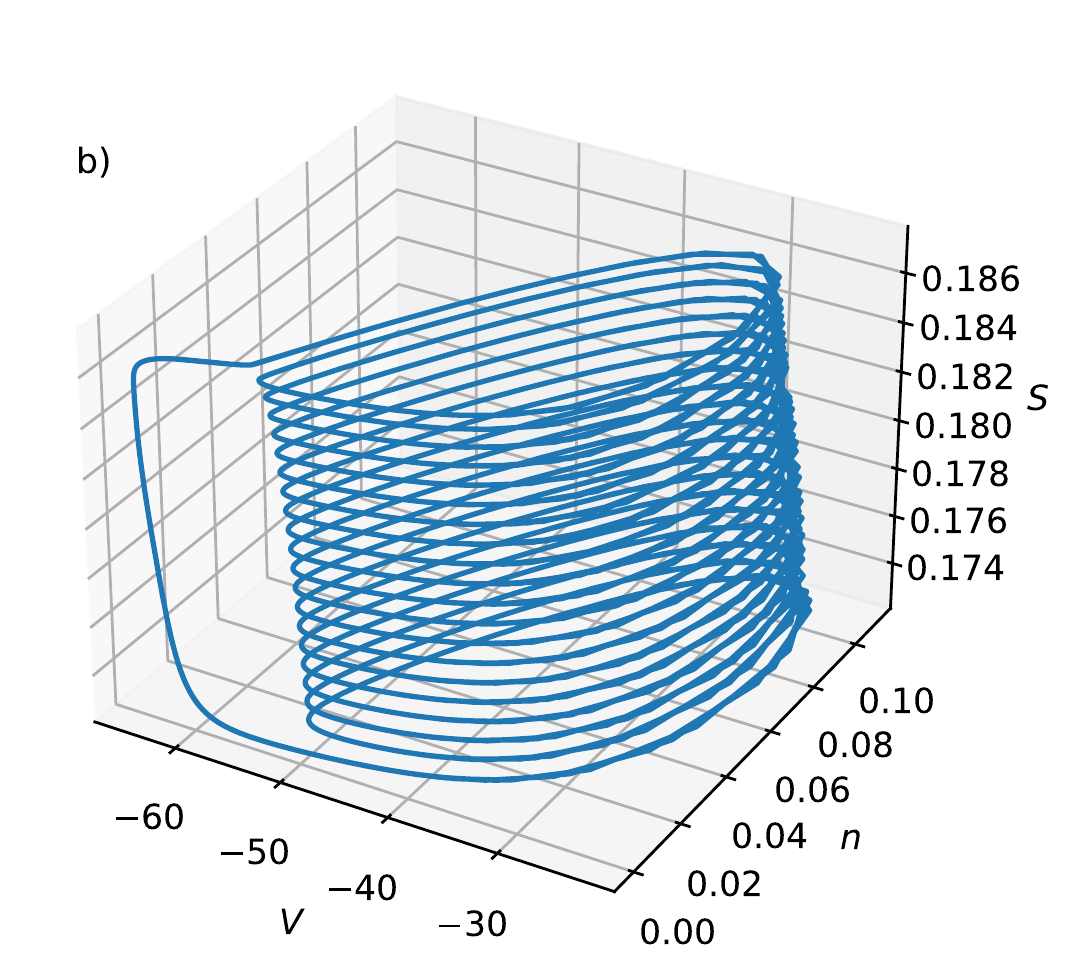} 
    \end{tabular}
  \end{center}
  \caption{\label{fig:attr3d}(a) Three-dimensional plot of the bursting
    attractor of the original system~\eqref{eq:sys}. Parameters are as in
    Fig.~\ref{fig:tser}(a,b,c). (b) Attractor of the corresponding neural
    network map~\eqref{eq:netw_model} trained to recover the original system in
    panel (a).}
\end{figure}

We will consider the system~\eqref{eq:sys} for $V_S$ varying within the range
$[-40, -30]$ while all other numerical values of parameters are listed in
Tab.~\ref{tab:param}. Parameter $g_{K2}$ is responsible for switching the
modification that introduces the bistability as discussed above. Setting
$g_{K2}=0$ we obtain the original system. This system has an unstable fixed
point. For example for $V_S=-36$ the fixed point is $V=-49.897$,
$n=2.3452\times 10^{-3}$, and $S=0.19946$. Its instability is indicated by the
positive largest eigenvalue: $\mu_1=20.772$, $\mu_2=9.0834\times 10^{-2}$,
$\mu_3=-42.884$.  Dynamics of the system~\eqref{eq:sys} in this case is
illustrated in Fig.~\ref{fig:tser}. Panels (a,b,c) represent busting
oscillations at $V_s=-36$. One can see that the variables $V$ and $n$ vary fast
while $S$ is slow. Figure~\ref{fig:attr3d}(a) shows the corresponding
three-dimensional phase portrait of the bursting attractor. When $V_S$ gets
larger the bifurcation occurs and after that the system demonstrates periodic
spiking, see Fig.~\ref{fig:tser} (d,e,f) plotted at $V_S=-33$.

\begin{figure}
  \begin{center}
    \includegraphics[width=0.99\columnwidth]{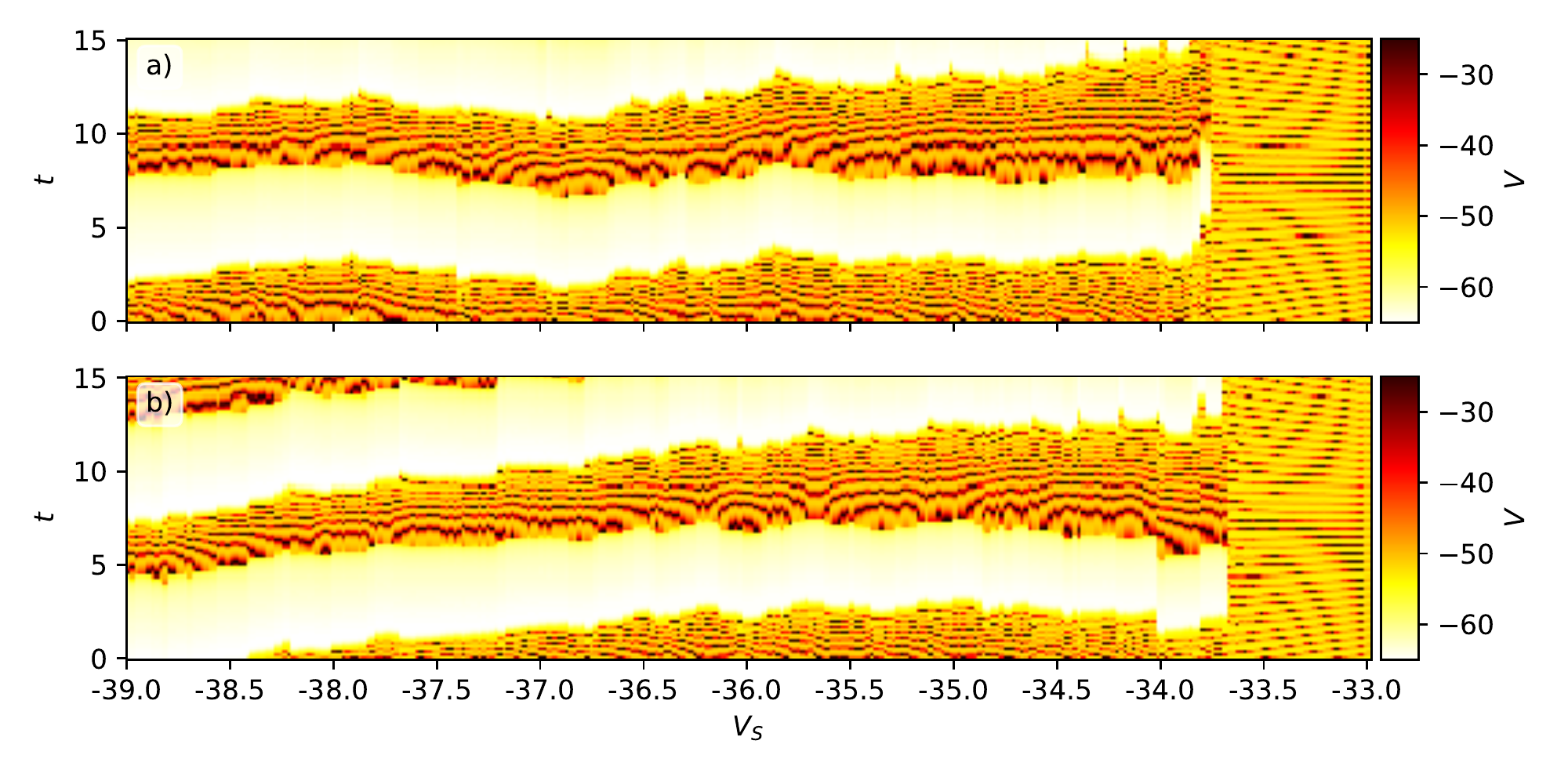}
  \end{center}
  \caption{\label{fig:bif_diag}(a) Bifurcation diagram of the original
    system~\eqref{eq:sys} at $g_{K2}=0$, i.e., without the bistability. Time
    dependence of $V$ is shown along the vertical axis with color shades, the
    darker the higher. Its modification with $V_S$ is demonstrated along the
    horizontal axis. (b) Bifurcation diagram computed with the corresponding
    neural network map~\eqref{eq:netw_model} trained to recover the original
    system.}
\end{figure}

Figure~\ref{fig:bif_diag}(a) shows how dynamics of the original
system~\eqref{eq:sys} at $g_{K2}=0$ changes as $V_S$ is varied. We will call it
bifurcation diagram. Here $V_S$ varies along the horizontal axis, time goes
vertically and shades of color indicate values of dynamical variable $V$
recorded after omitting transients (100 time units): darker color corresponds to
higher $V$. The data for this figure are computed while moving along $V_S$ from
left to right. Without a special taking care the bursting clusters at successive
parameter steps emerge shifted in time so that the neighboring patterns in the
plot are not adjusted and the whole picture is obscured. To improve the clarity
we compute each new solution (after dropping out the transients) over the
doubled time interval, i.e. for 30 time units instead of the used in the figure
15. Then within this longer solution we seek for a window that has the smallest
Euclidean distance with the solution already selected for the previous step. As
a result in Fig.~\ref{fig:bif_diag}(a) the area of bursting oscillations is
shown as two stripes that disappears at the bifurcation point
$V_S\approx -33.73$. To the right of this point high frequency spiking
oscillations appear that are represented in the diagram as a more or less
uniform texture.

\begin{figure}
  \begin{center}
    \begin{tabular}{cc}
      \includegraphics[width=0.48\columnwidth]{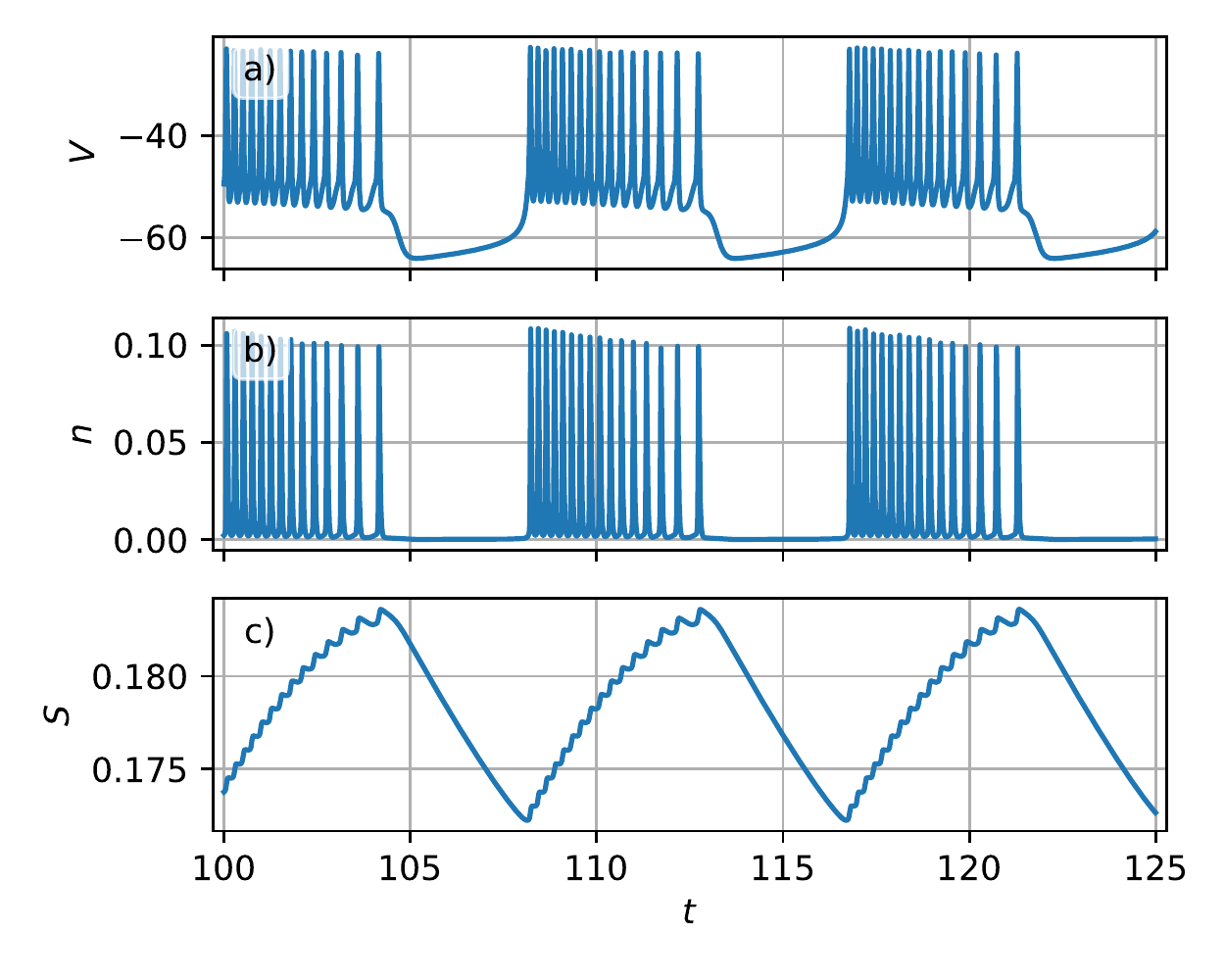} &
      \includegraphics[width=0.48\columnwidth]{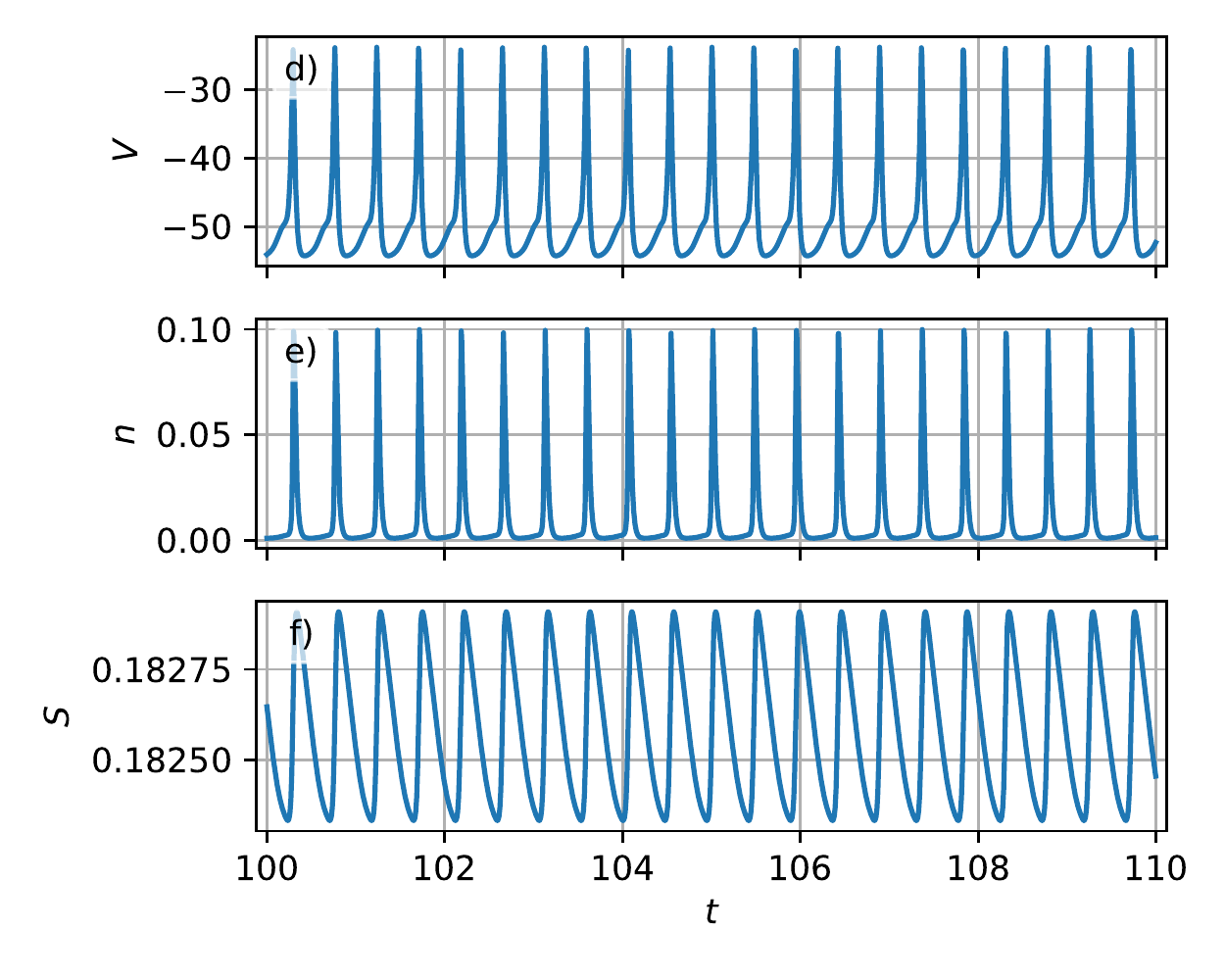}
    \end{tabular}
  \end{center}
  \caption{\label{fig:tser_mdf}Time series of the modified system~\eqref{eq:sys}
    at $g_{K2}=0.12$. (a,b,c) Bursting at $V_S=-36$. (d,e,f) Spiking at
    $V_S=-33$.}
\end{figure}

As discussed above the modification is engaged at nonzero $g_{K2}$: within a
certain range of $V_S$ the fixed point becomes stable, and it coexists with the
stable bursting attractor. At $g_{K2}=0.12$ and $V_S=-36$ the fixed point is
$V=-50.636$, $n=2.0560\times 10^{-3}$, $S=0.18792$, and its stability is
indicated by the negative largest eigenvalue: $\mu_1=-0.15927$, $\mu_2=-19.521$,
$\mu_3=-38.785$. Dynamics of the modified system is illustrated in
Fig.~\ref{fig:tser_mdf}. Bursting oscillations at $V_S=-36$ are shown in
Fig.~\ref{fig:tser_mdf}(a,b,c) and spiking oscillations at $V_S=-33$ are in
Fig.~\ref{fig:tser_mdf}(d,e,f). Observe that that the bursting and spiking of
the modified bistable system are visually almost indistinguishable with those
for the original system, cf. Fig.\ref{fig:tser}. The three-dimensional view of
the bursting attractor in this case also looks the same as for the original
system in Fig.~\ref{fig:attr3d}(a) and is not shown.

\begin{figure}
  \begin{center}
    \includegraphics[width=0.99\columnwidth]{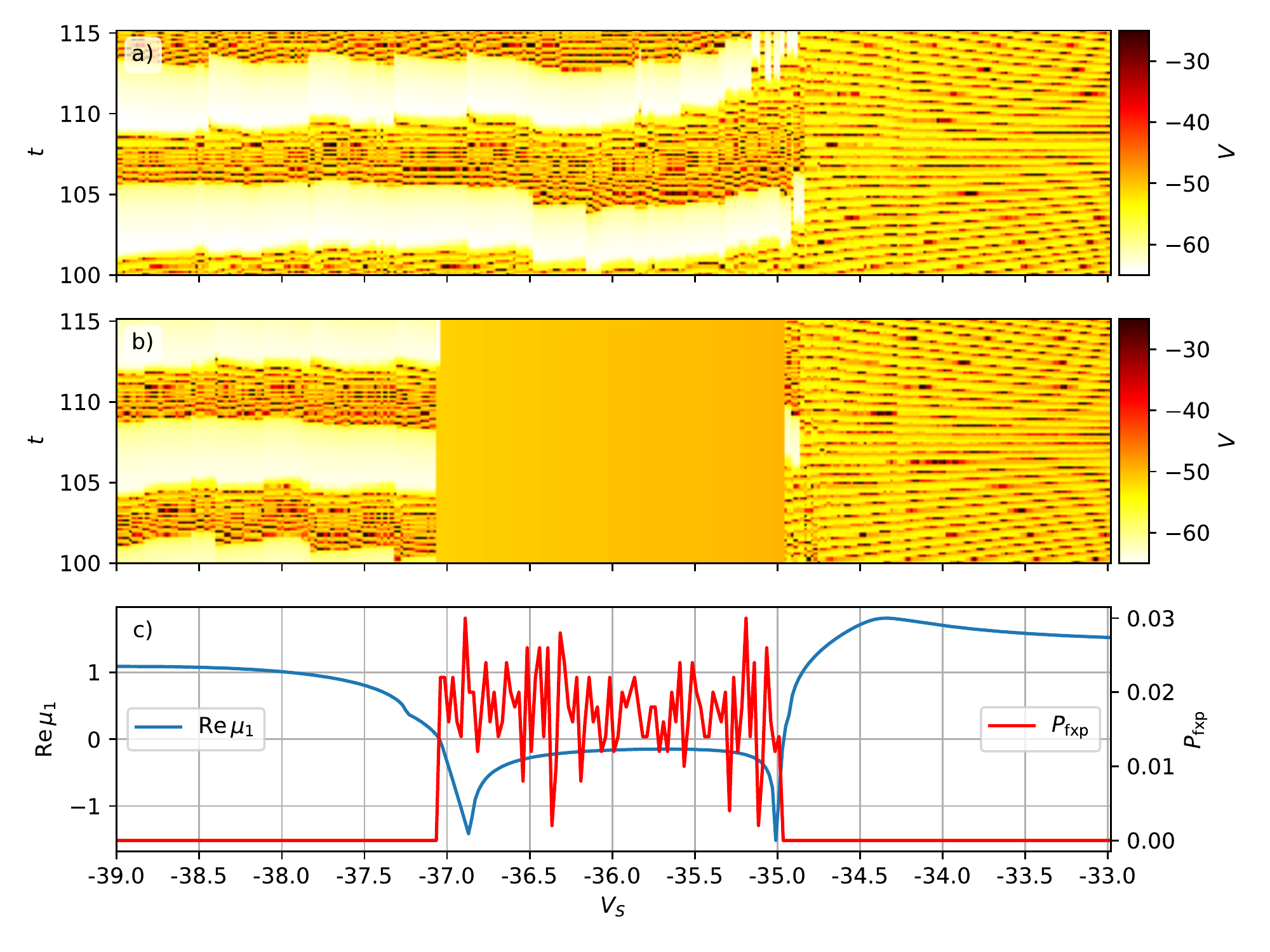}
  \end{center}
  \caption{\label{fig:bif_diag_mdf}Bifurcation diagrams of the modified
    system~\eqref{eq:sys} at $g_{K2}=0.12$. Panels (a) and (b) are computed with
    inheritance for two branches of solutions. Smooth area in the middle of the
    panel (b) indicate the stable fixed point. Panel (c) shows the largest real
    part of the eigenvalues of the fixed point (blue color). To improve the
    visibility positive values are plotted as $\log_{10}(\Re\mu_1+1)$ while the
    negative values are plotted as they are. Red curve in the panel (c) shows
    the relative frequency $P_\mathrm{fxp}$ of the stable fixed point
    implementation for random initial conditions, scale for it in right.}
\end{figure}

Coexistence of bursting and silent states in the modified system~\eqref{eq:sys}
at $g_{K2}$ is illustrated in the bifurcation diagrams in
Fig.~\ref{fig:bif_diag_mdf}. Axes and color encoding are the same as in
Fig.~\ref{fig:bif_diag}. To compute data for Fig.~\ref{fig:bif_diag_mdf}(a) we
take the starting point at $V_S=-36$ sufficiently far from the fixed point,
compute a solution and then find solutions moving to the left along the
parameter axis with the inheritance: starting point for the next parameter value
is taken as the solution from the previous point. Solutions are adjusted in time
to provide more clear picture in the same way as for
Fig.~\ref{fig:bif_diag}. Then we move in the same way from $V_S=-36$ to the
right. Figure~\ref{fig:bif_diag_mdf}(b) is computed in the same way, but the
initial staring point is taken close to the fixed point. We see the presence of
the bistability area in the middle parts of the plots: in
Fig.~\ref{fig:bif_diag_mdf} solutions initiated far from the fixed point
demonstrate bursting while trajectories started in the vicinity of the fixed
point converge to it. Figure~\ref{fig:bif_diag_mdf}(c), blue curve, shows how
the largest real part of the eigenvalues $\Re\mu_1$ of the fixed point depends
on the parameter. We see that in accordance with the observations in
Fig.~\ref{fig:bif_diag_mdf}(a,b) the fixed point has the negative eigenvalues in
the middle area so that it is stable and coexists with the bursting attractor.

The stable fixed point has a very small basin of attraction and, consequently,
there is very low probability that the system will arrive at it from random
initial conditions. To demonstrate it we consider a cube $V\in [-70,-10]$,
$n \in [0, 0.12]$, and $S \in [0.17, 0.2]$ that is large enough to cover both
the bursting attractor and the fixed point for any $V_S$ within the considered
range. For each $V_S$ we start 500 trajectories from random points within this
cube and count how many of them arrive at the fixed point. The corresponding
relative frequency $P_{\text{fxp}}$ is shown in Fig.~\ref{fig:bif_diag_mdf}(c)
with the red curve. We observe that $P_{\text{fxp}}$ oscillates near one - three
percents.

\section{\label{sec:netw}Neural network map}

Now we turn to reproducing the dynamics of the original and modified versions of
the system~\eqref{eq:sys} using an artificial neural network. Our goal is to
create the network that can be described as an explicit recurrent map of the
form $u(t+\Delta t)=F(u(t), p, w)$, where $\Delta t$ is a constant time step,
$u$ is a vector of dynamical variables, $p$ is a vector of control parameters,
and $w$ is a collection of inner parameters that are tuned in course of the
network training to obtain the desirable behavior.

Although we build a network that will operate as a recurrent map, it is created
and trained as a feed forward network. We avoid using a recurrent network
architecture since simple recurrent cells have the well known problems like
exploding gradients, i.e., instabilities in course of training, and fast
decaying memory~\cite{Goodfellow2017}. These problems have been solved for
more effective recurrent cells like LSTM or GRU, but they have complicated inner
structure that discourages their subsequent consideration as dynamical systems
prone to theoretical analysis. Also we do not consider an approach based on
reservoir computing since the state space dimension of a system on the basis of
the reservoir is much higher then the dimension of the modeled
system~\cite{Kong2021a,Kong2021b}.

In our previous work~\cite{NNDyn21} we considered a neural network model for
nonlinear dynamical systems in a form of the recurrent map
$u_{n+1}=u_{n}+\sigma(u_n A_0 + p B_0 + a_0) A_1 + a_1$, where $p$ is a vector
of control parameters, $A_{0,1}$ and $B_0$ are matrices, $a_{0,1}$ are vectors
and $\sigma(x)=(1+e^{-x})^{-1}$ is a sigmoidal function. High quality of
reproduction has been achieved for Lorenz and R\"{o}ssler systems as well as for
Hindmarsh-Rose model of neuronal activity. However the system~\eqref{eq:sys} is
found to require a more sophisticated approach. We address it to the presence of
components with very different time scales: $V$ and $n$ vary very fast and $S$
is slow, see Figs.~\ref{fig:tser}(a,b,c) and \ref{fig:tser_mdf}(a,b,c). It is
known that ODEs with such property requires special stiff solvers. So the
modeling of this dynamics with neural networks also requires an appropriate
network architecture.

Analysis of the problem reveals that the key point for modeling stiff dynamics
is to reproduce each dynamical variable separately. The following subnetwork
structure is suggested for the $i$th variable:
\begin{equation}
  \label{eq:netw_model}
  u_i(t+\Delta t) = u_i(t) +
  f(u_i(t) a_i + \alpha_i +g(u(t) A_i + p B_i + \beta_i)) b_i + \gamma_i.
\end{equation}
Here $u_i(t)$ is a scalar variable, either $V(t)$, $n(t)$ or $S(t)$, and $u(t)$
is a full vector of these dynamical variables. As usual for neural network
context we consider it is as a row vector. Its dimension is $D_u=3$. $p$ is a
row vector of control parameters of dimension $D_p$. We consider only one
varying parameter $V_S$ so that $D_p=1$. $a_i$, $\alpha_i$ and $\beta_i$ are row
vectors of dimension $N_h$, while $b_i$ is a column vector of dimension
$N_h$. $\gamma_i$ is a scalar. $A_i$ is $D_u$ by $N_h$ matrix. Its $i$th row is
assumed to contain zeros only. $B_i$ is $D_p$ by $N_h$ matrix. Functions
$f(\cdot)$ and $g(\cdot)$ are scalar and are assumed to be applied in
element-wise manner to vector elements. Both the functions $f(\cdot)$ and
$g(\cdot)$ and $N_h$ can in principle be chosen different for each subsystem but
for the sake of simplicity we do not apply a thorough hyperparameter
optimization to our network and take $N_h=100$ and hyperbolic tangent
$f(x)=g(x)=\tanh(x)$ that work well.  Table~\ref{tab:netw_param} contains the
whole network hyperparameters list.

\begin{table}
  \caption{\label{tab:netw_param}Hyperparameters of the network and the
    datasets}
  \begin{center}
  \begin{tabular}{c}
    \hline
    $f(x)=g(x)=\tanh(x)$, $N_h=100$, \\
    % S is chunk size, C is number of chunks
    $S=1000$, $C=1000$, $N_{\text{ds}}=S\times C=10^6$, \\
    $p_{\text{split}}=0.2$, $N_{\text{batch}}=1000$, \\ 
    $t_0=200$, $\Delta t = 0.005$. \\
    \hline
  \end{tabular}
  \end{center}
\end{table}

The idea behind forming the structure of the map~\eqref{eq:netw_model} is as
follows. In view of the theorems treating neural networks as universal
approximators~\cite{Kolmogorov56,Kolmogorov57,Arnold57, Cybenko1989,Haykin2009}
the form of the network for a scalar variable can be
$u_i(t+\Delta t) = u_i(t) + f(u_i(t) a_i + \alpha_i) b_i + \gamma_i$. This is
two layer network with one hidden layer of dimension $N_h$ whose state is given
by the output of $f(\cdot)$. Provided that $f(\cdot)$ is sigmoidal, e.g., the
hyperbolic tangent, and $N_h$ is sufficiently large these two layers are enough
to approximate almost any function with a required precision. In particular it
is able to approximate the map that reproduces dynamics of $u_i$. Yet this
structure does not take into account the vector of control parameters $p$ and
other dynamical variables. We suggest to inject them as the output of another
layer $h_i= g(u(t) A_i + p B_i + \beta_i)$. As already mentioned above the
matrix $A_i$ has zeros on its $i$th row so that the influence of the $i$th
variable onto $h_i$ is eliminated. This is done to reduce complexity of the
resulting recurrent map. Thus the full formula reads
$u_i(t+\Delta t) = u_i(t) + f(u_i(t) a_i + \alpha_i + h_i) b_i + \gamma_i$ that
exactly correspond to Eq.~\eqref{eq:netw_model}.

Thus the map~\eqref{eq:netw_model} is created as neural network whose weight
coefficients are collected in the following matrices and vectors
\begin{equation}
  \label{eq:all_weights}
  w = \{
  a_i, \alpha_i, A_i, B_i, \beta_i, b_i, \gamma_i\;|\; i=1,2,3
  \}.
\end{equation}
They have to be tuned in course of training.

Actual implementation of the network includes the following steps. First of all
we need to take into account that contemporary neural network frameworks as well
as their theoretical background assume that the datasets on which they operate
are standardized, i.e. they have zero mean and unit standard deviation. Training
on non standardized data will be highly ineffective. Thus the datasets engaged
in training of our networks are standardized before training as follows:
\begin{equation}
  \label{eq:normalize}
    u \to (u - m_u) / s_u, \;  p \to (p - m_p) / s_p,
\end{equation}
were $m_u$ and $m_p$ are vectors of mean vales of $u$ and $p$, respectively, and
$s_u$ and $s_p$ are the corresponding standard deviations. The element-wise
operations are assumed. All training routines are done for the rescaled
data. When the network is in use we first rescale the initial states and the
parameters according to Eqs.~\eqref{eq:normalize} then perform computations and
finally apply the inverse of these transforms before showing the results.

\newcommand{\layDense}{\mathop{\mathrm{Dense}}}
\newcommand{\layInput}{\mathop{\mathrm{Input}}}
\newcommand{\layTake}{\mathop{\mathrm{Take}}}
\newcommand{\laySans}{\mathop{\mathrm{Sans}}}
\newcommand{\sansU}[1]{u_{\neg\,#1}}

The neural network uses the following operators. Operator $\layTake(u,i)$
returns $i$th elements of a vector $u$, and $\laySans(u,i)$ returns the vector
without its $i$th element. Operator $\layDense(x, W, b) = x W + b$ represents a
standard dense layer that mathematically corresponds to an affine transformation
of the vector with matrix $W$ and vector $b$. Notice that in our notation an
activation function is not built-in into the dense layer. It will be applied
separately.  There is the matrix $A_i$ in~\eqref{eq:netw_model} whose $i$th row
contains zeros. For the actual network training and operation we use a matrix
$A'_i$ whose $i$th row is absent that fulfills the vector $\sansU{i}$ whose
$i$th is also removed. The square brackets $[\cdot,\cdot]$ are used to indicate
concatenation of two vectors or matrices.

The neural network implementing the map~\eqref{eq:netw_model} can be described
as follows. It has two input vectors, $p=(V_S)$ (actually the vector $p$ is
one-dimensional) and $u=(V, n, S)$, Eq.~\eqref{eq:netw_up}. First $u$ is split
into a scalar $u_i$ and a vector $\sansU{i}$ (its $i$th element is absent), see
Eq.~\eqref{eq:netw_split}. The layer~\eqref{eq:netw_lay1} takes $\sansU{i}$ and
$p$ and computes $N_h$ dimensional vector $h_i$. This vector assembles the
influence of the control parameters and of all but $i$th dynamical
variables. This vector is added to the output of the next dense layer that
process the $i$th component itself, Eq.~\eqref{eq:netw_lay2}. The result of this
transformation is $N_h$ dimensional vector $q_i$ that finally goes to the third
layer \eqref{eq:netw_lay3} whose output is the increment to the initial value
$u_i$ to obtain a value at the next time step $v'_i$ that is the network output.
Doing in the same way for all $i=1,2,3$ we form a full output vector $v'$.

\begin{gather}
  \label{eq:netw_up}
  p = \layInput(),\; u = \layInput(),\\
  \label{eq:netw_split}
  u_i = \layTake(u,i),\; \sansU{i} = \laySans(u, i),\\
  \label{eq:netw_lay1}
  h_i=g(\layDense([\sansU{i}, p], [A'_i, B_i], \beta_i)), \\
  \label{eq:netw_lay2}
  q_i=f(\layDense(u_i,a_i, \alpha_i) + h_i), \\
  \label{eq:netw_lay3}
  v'_i = u_i + \layDense(q_i, b_i, \gamma_i)
\end{gather}

The network is trained as a feed forward network. The dataset is prepared as
follows: choose the time step $\Delta t$, see Tab.~\ref{tab:netw_param}, and
prepare the dataset as a collection of $N_{\text{ds}}$ records ($p$, $u$, $v$),
where $u$ is an initial point for Eqs~\eqref{eq:sys} solved at the control
parameters $p$ during time interval $\Delta t$ to obtain $v$. This dataset is
split into training and validation parts with
$N_{\text{ds}}(1-p_{\text{split}})$ records in the training part and
$N_{\text{ds}}\,p_{\text{split}}$ for the validation, see
Tab.~\ref{tab:netw_param} for $p_{\text{split}}$ value. The dataset preparation
will be discussed in more detail below.

In course of training a pair of vectors $p$ and $u$ are fed at the input, the
network response vector $v'$ is computed, and it is compared with the correct
value $v$ from the dataset. The loss function for the training is mean squared
error:
\begin{equation}
  \label{eq:netw_loss}
  L = \|v - v'\|^2.
\end{equation}
The goal of the training is to minimize $L$ on the dataset due to tuning its
parameters $w$, Eq.~\eqref{eq:all_weights}. In the very beginning the parameters
are initialized at random. The whole training dataset is shuffled and split into
batches each of $N_{\text{batch}}$ records, see Tab.~\ref{tab:netw_param}. Each
batch is sent to the network and the loss function~\eqref{eq:netw_loss} is
computed for the batch as well as its gradient $\nabla_w L$ with respect to
$w$. Then it is used in a gradient descent step to compute updates to $w$. The
simplest version of the gradient descent step reads
\begin{equation}
  \label{eq:grad_descent}
  w \rightarrow w - \gamma \nabla_w L
\end{equation}
where the step size scale $\gamma$ is a small parameter controlling the
convergence. In actual computations instead of the simplest one a more
sophisticated method is used called Adam~\cite{Kingma2014}. The difference
is that the step size scale $\gamma$ is not a constant, but is tuned according
to the accumulated gradients on the previous steps. When all batches are
processed this is called an epoch. The training dataset is shuffled again and a
new epoch of training starts. Quality of the training is estimated after each
epoch by feeding the network with the validation data and computing the loss
function without updating $w$. We perform $2000$ epochs of training and after
that the mean squared error on the validation data decay down to the level
$10^{-7}$.

Two networks are created as described above, for the original as well as for the
modified system~\eqref{eq:sys}. Having the same structures described by
Eq.~\eqref{eq:netw_model} the networks differ by sets of parameters $w$ obtained
after training on two sets of data computed using Eqs.~\eqref{eq:netw_model}.

To create the datasets we fix ODEs~\eqref{eq:sys} parameters either for the
original, $g_{K2}=0$, or for the modified system, $g_{K2}=0.12$, see
Tab.~\ref{tab:param} and sample $C$ parameter vectors $p=(V_S)$ from the uniform
distribution on the range $[-40, -30]$. For each parameter value a random
initial point is generated $u(0)=(V(0),n(0),S(0))$ sufficiently far from the
fixed point to make sure that the trajectory even in the bistability case will
not go to it.  Then Eqs.~\eqref{eq:sys} are solved starting from this
point. After a transient $t_0$, see Tab.~\ref{tab:netw_param}, the system
arrives at bursting or spiking attractor and its points are recorded with time
step $\Delta t$ in a row: $u(t_0)$, $u(t_0+\Delta t)$, $u(t_0+2\Delta t)$,
$u(t_0+3\Delta t)$, \dots Totally $S+1$ points are recorded. After that $S$ data
records are composed as ($p$, $u(t_0)$, $u(t_0+\Delta t)$), ($p$,
$u(t_0+\Delta t)$, $u(t_0+2\Delta t)$), ($p$, $u(t_0+2\Delta t)$,
$u(t_0+3\Delta t)$) and so on.  Notice that when training the trajectory chunks
are shuffled so that each batch is a random sample of pairs of points, not the
whole cuts of trajectories.

Altogether the dataset is formed by $C$ chunks of trajectories. Each chunk
corresponds to a certain parameter value $V_S$ and includes $S$ trajectory
points sampled with the time step $\Delta t$ in a row on either bursting or
spiking attractors. Fixed point is never recorded to the dataset even in the
case of bistability when the fixed point is stable. Thus in course of training
the network never sees it.

\begin{figure}
  \begin{center}
    \includegraphics[width=0.7\columnwidth]{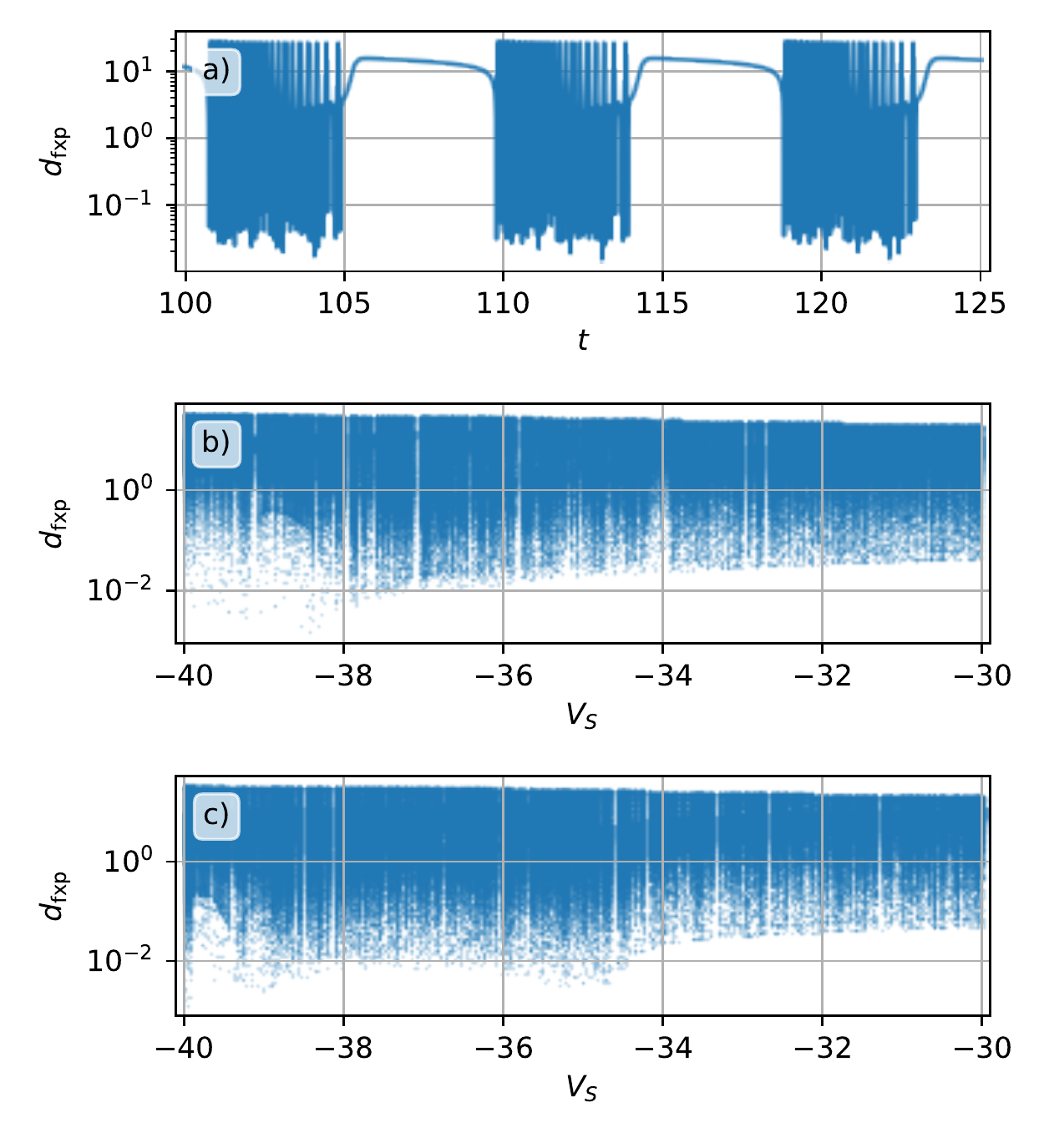}
  \end{center}
  \caption{\label{fig:fxp_dist}(a) Distance to the fixed point for a typical
    trajectory on the bursting attractor. (b) and (c) Distance to the fixed
    point for the dataset points used for training the neural network
    map~\eqref{eq:netw_model} for the original and the modified systems,
    respectively.}
\end{figure}

The structure of the dataset generated for training the networks are illustrated
in Fig.~\ref{fig:fxp_dist}. Figure~\ref{fig:fxp_dist}(a) represents a typical
example of time dependence of distance from the oscillating attractor and the
fixed point. We observe the fixed point is clearly separated from the
oscillating attractor so that the distance cannot be smaller then
$10^{-2}$. Figures~\ref{fig:fxp_dist}(b,c) show distances from the fixed point
to the dataset points for the original and the modified systems,
respectively. We see that the fixed point remains separated from the dataset for
the whole parameter range.

Since the neural network map~\eqref{eq:netw_model} has an explicit form we can
explicitly compute its Jacobian matrix. This matrix is required both for a
numerical routines finding the fixed point and for testing stability of the
fixed point. The Jacobian matrix is found by differentiating right hand side of
Eq.~\eqref{eq:netw_model} by $u_j$, where $j=1,2,3$. After straightforward
computations we can write the following expressions for the diagonal and
off-diagonal elements of the matrix: %
\newcommand{\opDiag}{\mathop{\mathrm{Diag}}}
\newcommand{\opRow}{\mathop{\mathrm{Row}}}
\begin{equation}
  \begin{gathered}
    \label{eq:netw_jac_elems}
    j_{ii} = 1 + a_i\opDiag f'(z_i) b_i, \\
    j_{ij} = \opRow(A_i, j) \opDiag g'(y_i) \opDiag f'(z_i) b_i,\\
    z_i = u_i a_i + \alpha_i +g(y_i), \\
    y_i = u A_i + p B_i + \beta_i
  \end{gathered}
\end{equation}
Here operator $\opDiag(x)$ creates a diagonal matrix whose diagonal consists of
elements of vector $x$. Operator $\opRow(X, i)$ returns a row vector whose
elements are taken as $i$th row of the matrix $X$.

\section{\label{sec:dynam}Dynamics of the neural network map}

\begin{figure}
  \begin{center}
    \begin{tabular}{cc}
      \includegraphics[width=0.48\columnwidth]{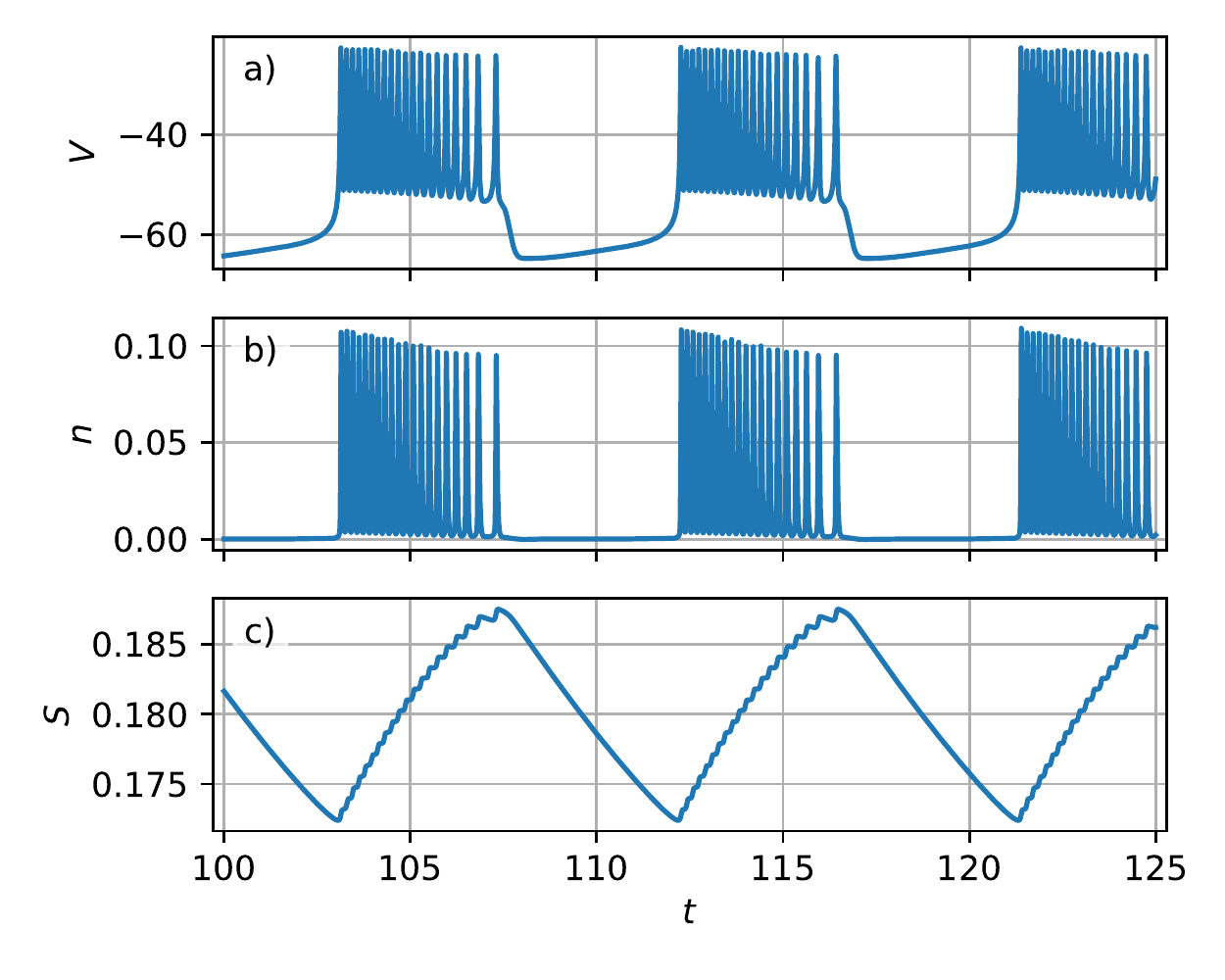} &
      \includegraphics[width=0.48\columnwidth]{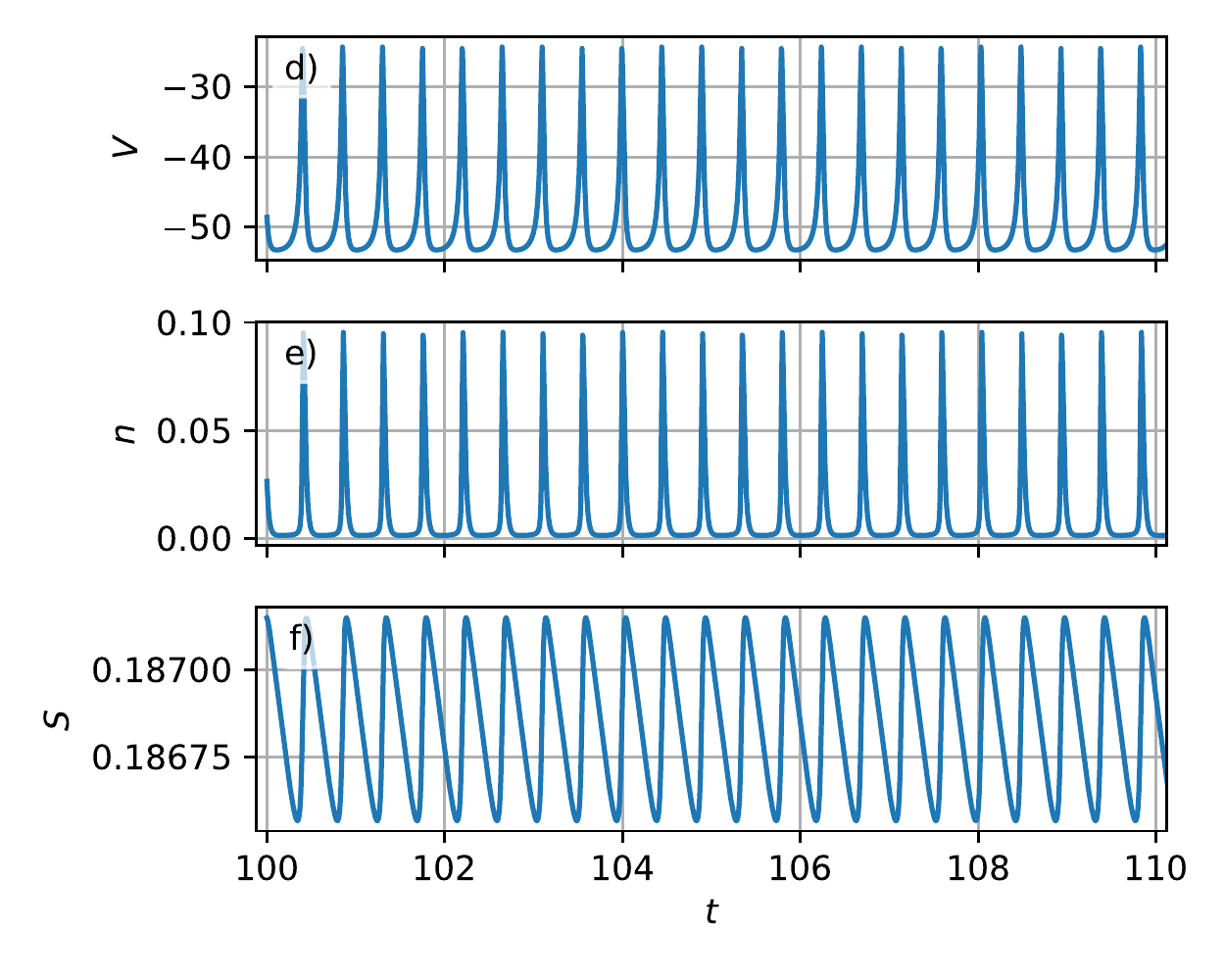}
    \end{tabular}
  \end{center}
  \caption{\label{fig:tser_ntw}Time series same as in Fig.\ref{fig:tser} for the
    neural network map~\eqref{eq:netw_model} trained to reproduce the original
    system.}
\end{figure}

\begin{figure}
  \begin{center}
    \begin{tabular}{cc}
      \includegraphics[width=0.48\columnwidth]{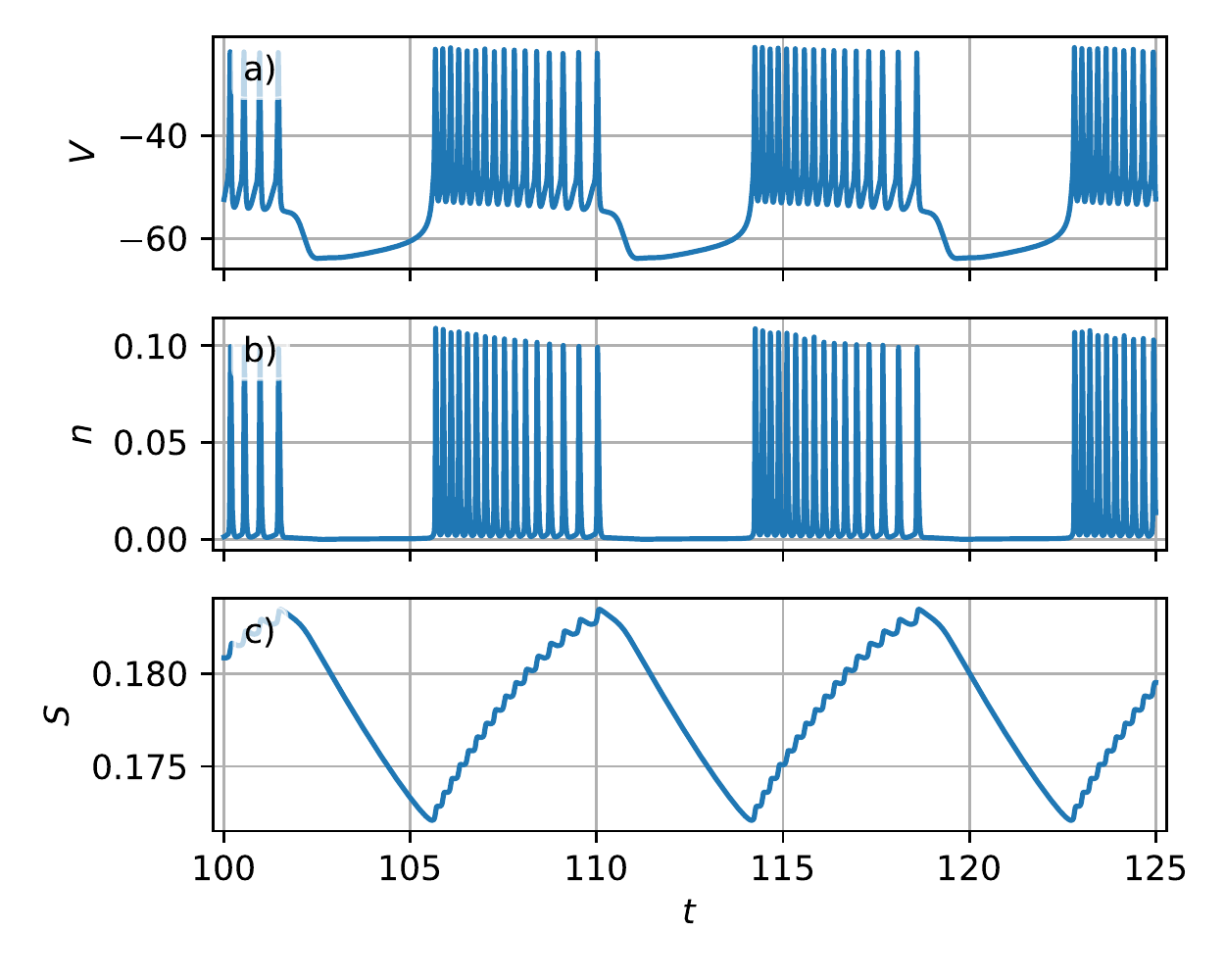} &
      \includegraphics[width=0.48\columnwidth]{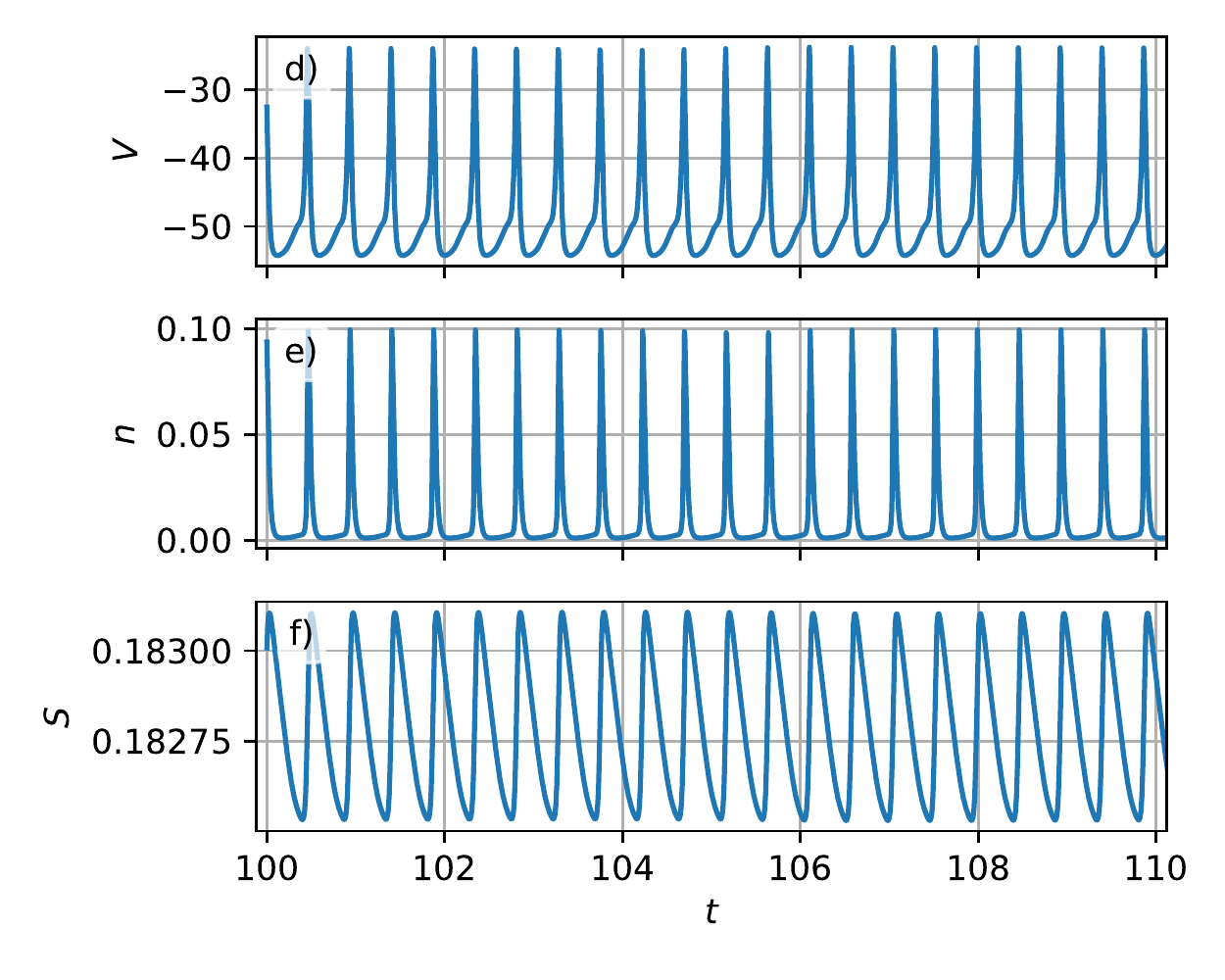}
    \end{tabular}
  \end{center}
  \caption{\label{fig:tser_ntw_mdf}Time series same as Fig.\ref{fig:tser_mdf}
    for the neural network map~\eqref{eq:netw_model}, modified system.}
\end{figure}

Using the described above architecture and datasets we have trained two neural
network maps of the form~\eqref{eq:netw_model} that reproduce dynamics of the
original and modified versions of the system~\eqref{eq:sys}. Let us first
demonstrate how these networks learn the attractors that were shown them during
the training. Figure~\ref{fig:tser_ntw} demonstrates solutions computed as
iteration of the neural network map~\eqref{eq:netw_model} for the original
system at the same parameters as ODEs in
Fig.~\ref{fig:tser}. Figure~\ref{fig:tser_ntw_mdf} represents solution computed
using the map~\eqref{eq:netw_model} for the modified system. The parameter
values are as in Fig.~\ref{fig:tser_mdf}.  Observe very high similarity of
trajectories of the neural network maps with the corresponding ODEs
solutions. Figure~\ref{fig:attr3d}(a,b) compares three dimensional views of the
bursting attractors for the original system~\eqref{eq:sys} and the corresponding
neural network map~\eqref{eq:netw_model}, respectively. Observe again that the
plots are very similar.

\begin{figure}
  \begin{center}
    \includegraphics[width=0.99\columnwidth]{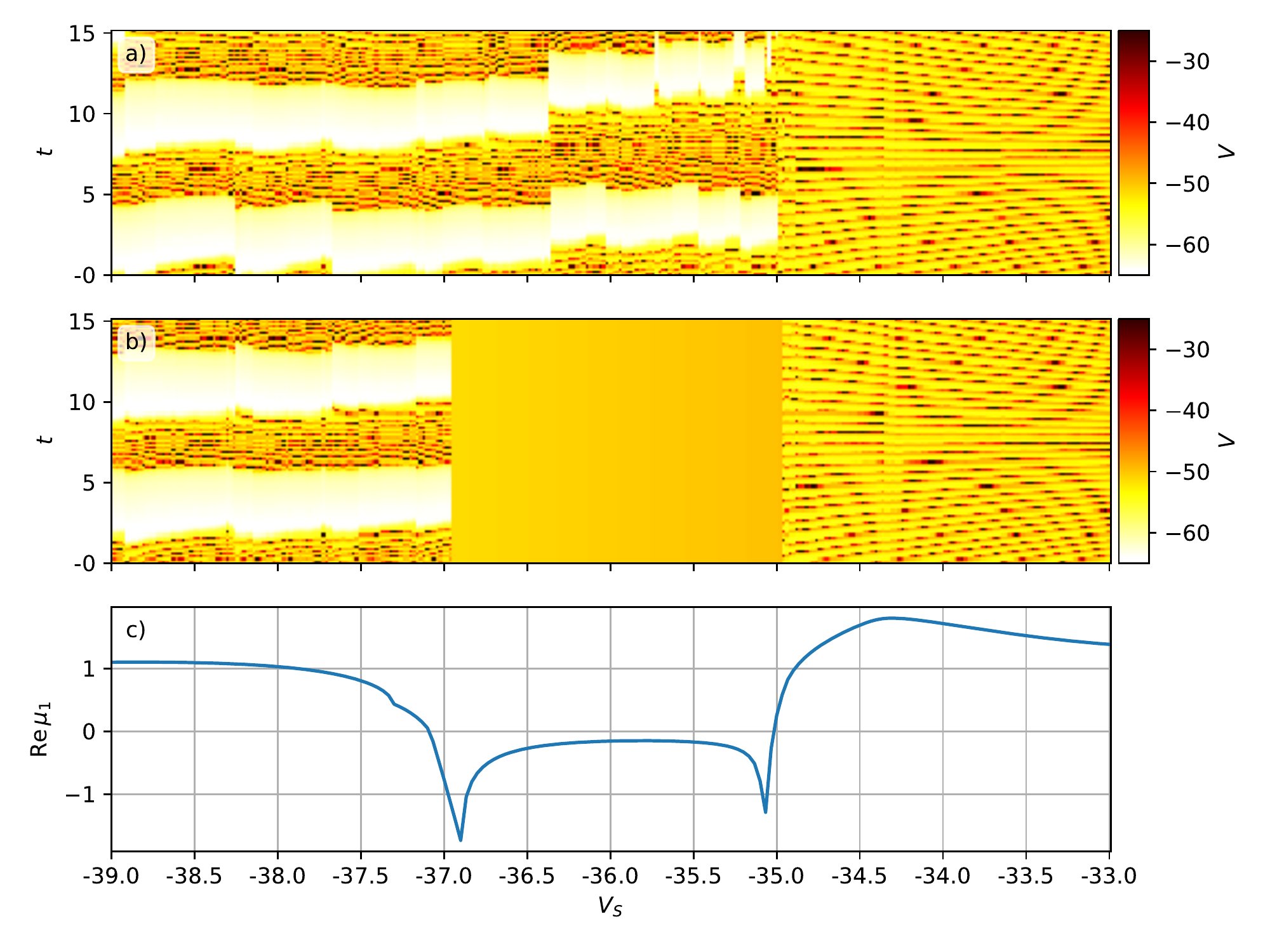}
  \end{center}
  \caption{\label{fig:bif_diag_mdf_netw}Bifurcation diagram for the neural
    network map~\eqref{eq:netw_model} trained for the modified system with the
    bistability. Panels are as in Fig.~\ref{fig:bif_diag_mdf}.}
\end{figure}

Bifurcation diagrams for the original system~\eqref{eq:sys} and for the
corresponding neural network map~\eqref{eq:netw_model} are shown in
Fig.~\ref{fig:bif_diag}(a,b), respectively. We see the very high similarity
again. The bifurcation diagrams for the modified system with bistability
computed with ODEs~\eqref{eq:sys} and with the neural network
map~\eqref{eq:netw_model} are compared in Figs.~\ref{fig:bif_diag_mdf}(a,b)
and~\ref{fig:bif_diag_mdf_netw}(a,b), respectively. We observe again a
remarkable correspondence.

\begin{figure}
 \begin{center}
  \includegraphics[width=0.99\columnwidth]{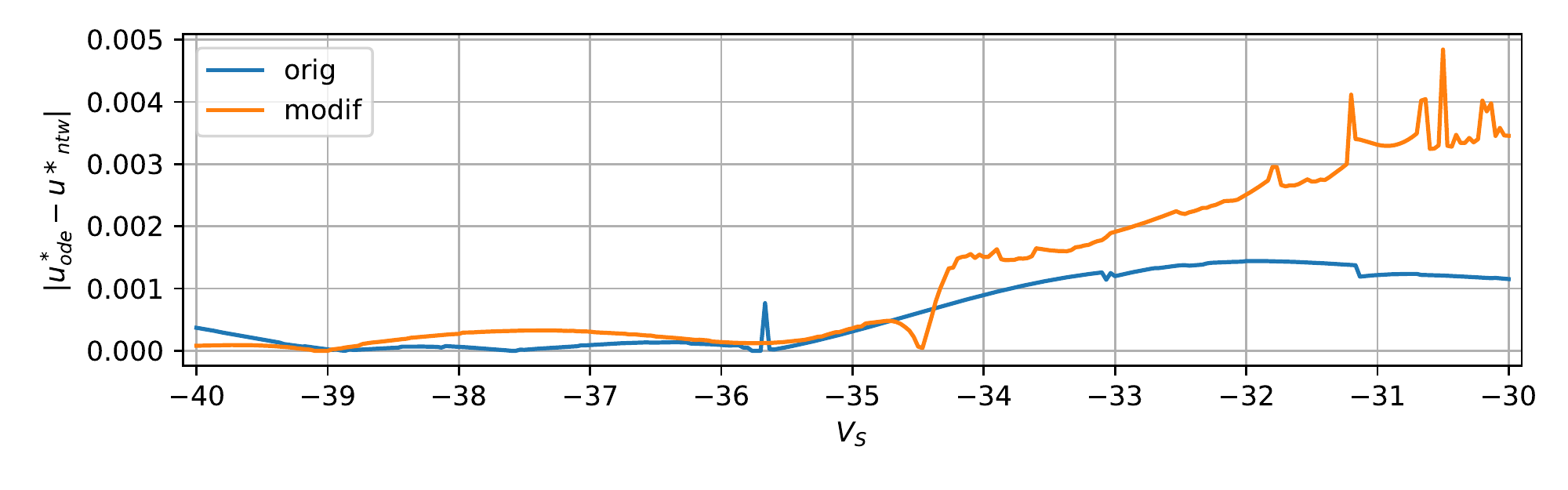}
 \end{center}
 \caption{\label{fig:fxp_err}Distance between the fixed point of
   ODEs~\eqref{eq:sys} and the neural network map~\eqref{eq:netw_model} for the
   original and the modified systems.}
\end{figure}

However the most interesting is that the neural network
map~\eqref{eq:netw_model} is also able to discover the fixed point of the
system~\eqref{eq:sys} that was never showed to it in the course of training.
Since this map as well as its Jacobian matrix~\eqref{eq:netw_jac_elems} are
known explicitly, we can use the standard numerical routines, e.g., Newton
method, to compute the fixed point of this map with high precision and compare
it with the fixed point of ODEs~\eqref{eq:sys}. The result of this comparison is
shown in Fig.~\ref{fig:fxp_err} that represents Euclidean distance between the
fixed points of ODEs \eqref{eq:sys} and the neural network
map~\eqref{eq:netw_model} for the original system as well as for the modified
system with bistability. Taking into account a large scale of the variable $V$,
see for estimation Figs.~\ref{fig:tser}(a) and \ref{fig:tser_mdf}(a), we can say
that the error in the fixed point reproduction of the order $10^{-3}$ can be
treated as a very small.

The ability of the network to discover the fixed point indicates that the
training data describing oscillating attractor implicitly contains sufficient
information about the fixed point and the network in course of training reveals
it.

The created networks not only find the fixed point location but also correctly
reveal its stability properties. Figure~\ref{fig:bif_diag_mdf_netw}(a) shows
that the network correctly reproduces the bifurcation diagram for the
oscillating branch of the bistability regime. It is this branch that has been
showed to the network in training. But in Fig.~\ref{fig:bif_diag_mdf_netw}(b) we
observe that the second, silent branch is also discovered. This branch did not
appear in the training data and recovered by the network due to its effective
generalization of the training data. Observe that boundaries of the bistability
area are found with very high accuracy. It means that the information about
loosing stability of the fixed point is also discovered by the network. Another
interesting point is that the bursting attractors for the original and for the
modified systems are visually almost indistinguishable, compare
Figs.~\ref{fig:tser}(a,b,c) and \ref{fig:tser_mdf}(a,b,c). But nevertheless the
network is able to distinguish them in the course of training: using data
produced by the original system results in the neural network map with the
unstable fixed point while the training data for the modified system results in
the neural network map with the bistability.

Figure~\ref{fig:bif_diag_mdf_netw}(c) shows the largest real part of the
eigenvalue of the fixed point $\mu_1$. Similarly to
Fig.~\ref{fig:bif_diag_mdf}(c) to adjust scales of positive and negative values
of the eigenvalue the positive values are plotted in logarithmic scale. As
expected, the areas of stability of the fixed point in
Fig.~\ref{fig:bif_diag_mdf_netw}(b) correspond to negative values of its largest
eigenvalue. Observe high similarity of the curves for eigenvalues in
Figs.~\ref{fig:bif_diag_mdf}(c) and \ref{fig:bif_diag_mdf_netw}(c) computed for
ODE~\eqref{eq:sys} and for the neural network map~\eqref{eq:netw_model},
respectively.

\begin{figure}
 \begin{center}
  \includegraphics[width=0.99\columnwidth]{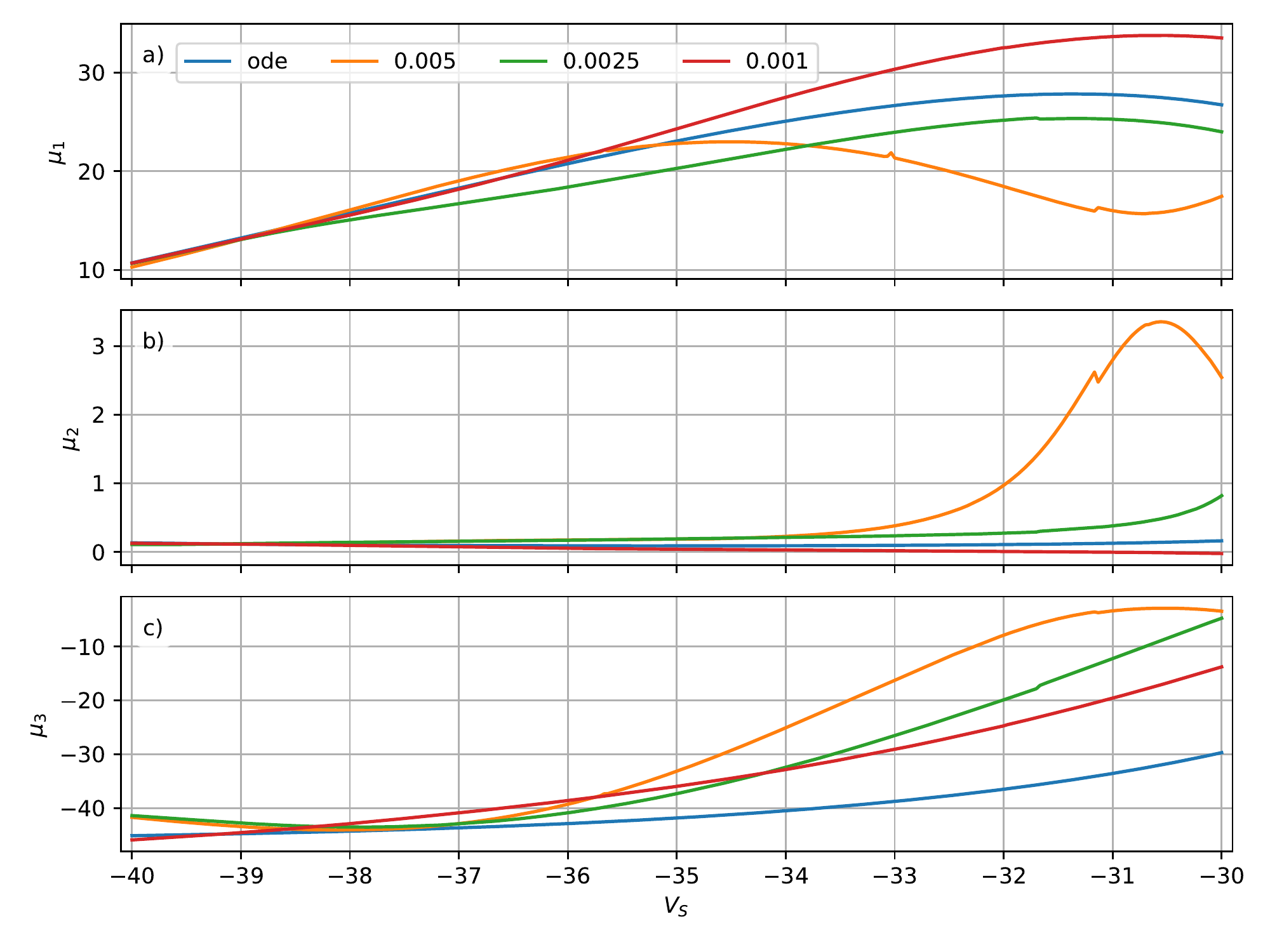}
 \end{center}
 \caption{\label{fig:eig_fxp}Eigenvalues of the fixed point of the
   ODEs~\eqref{eq:sys} and the neural network map~\eqref{eq:netw_model}, the
   original system. Three networks are considered trained for $\Delta t=0.005$,
   $0.0025$, and $0.001$.}
\end{figure}

\begin{figure}
 \begin{center}
  \includegraphics[width=0.99\columnwidth]{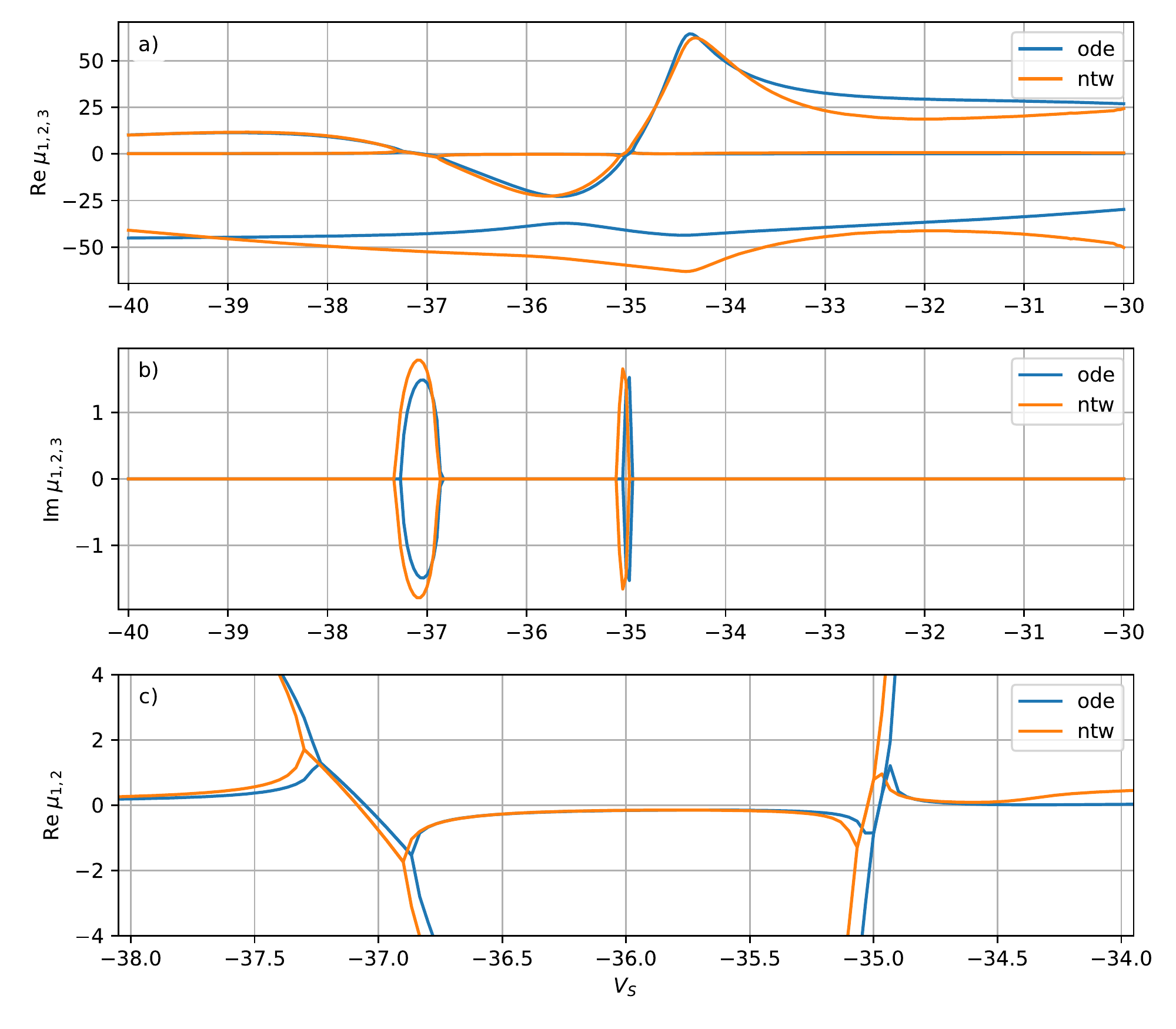}
 \end{center}
 \caption{\label{fig:mdf_eig_fxp}(a,b) Real and imaginary parts of the
   eigenvalues for the ODEs and the network, the modified system. (c) Real parts
   of the eigenvalues, enlarged area in the middle part of the panel (a).}
\end{figure}

In more details the eigenvalues of the fixed point are tested in
Figs.~\ref{fig:eig_fxp} and \ref{fig:mdf_eig_fxp}. Figure \ref{fig:eig_fxp}
shows the eigenvalues computed for the original system and for the corresponding
network map. The eigenvalues are real within the whole parameter range. Three
neural networks are represented. In addition to the main one at $\Delta t=0.005$
we also trained two networks with smaller time steps $\Delta t = 0.0025$ and
$\Delta t = 0.001$. Observe qualitative similarity of the eigenvalues computed
for the networks with the true values computed for ODEs: all of them are real,
two largest are positive for all cases. It means the trajectory leaving the
vicinity of the fixed point behave in the same ways in all cases: there are two
unstable directions and no rotation around it occurs since there are no
imaginary parts. Eigen values for the networks coincide very well with those for
ODEs in the left part of the figures and diverge in the right part. We
conjecture that the spiking attractor requires finer time resolution. In favor
of this assumption is the behavior of the eigenvalue curves computed for the
networks trained with smaller time steps. We see that decreasing the time step
results in better approximation of the true curve obtained for the ODEs.

The eigenvalues of the fixed point for the modified system with bistability is
shown in Fig.~\ref{fig:mdf_eig_fxp}. Panels (a) and (b) represent real and
imaginary parts, respectively. Observe remarkable correspondence between the
values computed for ODEs and for the network. Figure~\ref{fig:mdf_eig_fxp}(c)
shows two largest real parts of the eigenvalues near the area of bistability. We
see that the network reproduces sufficiently fine details: the area of the
negative values in the middle is bounded by segments with coinciding real
parts. As one can see in Fig.~\ref{fig:mdf_eig_fxp}(b) this corresponds to
complex conjugated eigenvalues. Then the eigenvalues again become real and
diverge from each other.

Thus we see that properties of the fixed points of the original and the modified
systems are remarkably discovered by the networks regardless of the fact that
they were never showed to the networks during training. It works well both for
the unstable fixed point and for the bistability case. For the latter it means
that it is enough to impose to the network only one branch of the bistable
solution and it reveals the regime of bistability and discovers the second branch.

\section{\label{sec:concl}Conclusion}

We created an artificial neural network that reproduces dynamics of
Hodgkin-Huxley-type model represented as a stiff ODE system with two fast and
one slow variables. Two versions (original and modified) of the model are
considered. For the considered parameter ranges the original version has
unstable fixed point and oscillating attractor that can be either bursting or
spiking. The modification introduces bistability such that an area in the
parameter space appears where the fixed point becomes stable and coexists with
the bursting attractor.

The created network operates as a recurrent map, i.e., reproduces the dynamics
as a discrete time system. This is trained as a feedforward network using
standard back propagation routine on trajectory cuts sampled at random parameter
values within a certain range. The network structure is developed to take into
account the stiffness of the modeled system. Due to different time scales it is
found to be effective to model each variable with a separate subnetwork. The
subnetworks have identical structures and contain three fully connected
layers. The first one injects the parameter values as well as non-modeled
variables. The output of this layer is added to a vector of weights of the
modeled variable and the result passes trough two another layers. Scalar outputs
of each subnetwork are concatenated to form a vector of dynamical variables at
new time step.

Although the network is trained only on oscillatory trajectory cuts, the
resulting recurrent map also acquires the fixed point whose position and even
the eigenvalues coincide well with the fixed point of the initial system. In
particular it means that when a bistable regime is modeled, the network being
fed by only one brunch of solutions discovers another brunch, never seen in the
course of training. It indicate the ability of the created network to perform
proper generalization of the training data.

The obtained results, i.e., the ability of the created artificial neural network
to reproduce dynamics including dynamical features never seen in the course of
training are able, as we see it, to trigger new approaches to complex dynamics
reconstruction problem. Potentially the methods of analysis can be developed
where neural network discovers previously unknown dynamical features of the
analyzed system. From the practical point of view reproducing dynamics with the
neural network can be considered as a sort of alternative method of numerical
modeling intended for use with contemporary parallel hard- and software. In
particular it is suitable for so called AI accelerators, a hardware dedicated to
deal with artificial neural networks.

\section*{Declaration of competing interest}

The authors declare that they have no known competing financial interests or
personal relationships that could have appeared to influence the work reported
in this paper.

\section*{Acknowledgement}

Work of PVK on theoretical formulation and numerical computations and work of
NVS and ERB on results analysis was supported by grant of Russian Science
Foundation No 20-71-10048, \\ https://rscf.ru/en/project/20-71-10048/

\bibliographystyle{elsarticle-num} 
\bibliography{nn-sherman}

\end{document}